\documentclass{elsart}

\usepackage[square,comma]{natbib}
\usepackage{graphicx}
\usepackage{pxfonts}

\usepackage{amssymb}

\journal{}

\begin{document}

\thispagestyle{empty}
\begin{Large}
\textbf{DEUTSCHES ELEKTRONEN-SYNCHROTRON}

\textbf{\large{in der HELMHOLTZ-GEMEINSCHAFT}\\}
\end{Large}

DESY 07-087

June 2007

\begin{eqnarray}
\nonumber &&\cr \nonumber && \cr \nonumber &&\cr
\end{eqnarray}
\begin{eqnarray}
\nonumber
\end{eqnarray}
\begin{center}
\begin{Large}
\textbf{Longitudinal impedance and wake from XFEL undulators.
Impact on current-enhanced SASE schemes}
\end{Large}
\begin{eqnarray}
\nonumber &&\cr \nonumber && \cr
\end{eqnarray}

\begin{large}
Gianluca Geloni, Evgeni Saldin, Evgeni Schneidmiller and Mikhail
Yurkov
\end{large}
\textsl{\\Deutsches Elektronen-Synchrotron DESY, Hamburg}
\begin{eqnarray}
\nonumber
\end{eqnarray}
\begin{eqnarray}
\nonumber
\end{eqnarray}
\begin{eqnarray}
\nonumber
\end{eqnarray}
ISSN 0418-9833
\begin{eqnarray}
\nonumber
\end{eqnarray}
\begin{large}
\textbf{NOTKESTRASSE 85 - 22607 HAMBURG}
\end{large}
\end{center}
\clearpage
\newpage

\begin{frontmatter}



\title{Longitudinal impedance and wake from XFEL undulators. Impact on current-enhanced SASE schemes}


\author{Gianluca Geloni,}
\author{Evgeni Saldin,}
\author{Evgeni Schneidmiller}
\author{and Mikhail Yurkov}

\address{Deutsches Elektronen-Synchrotron (DESY), Hamburg,
Germany}

\begin{abstract}
In this article we derive longitudinal impedance and wake function
for an undulator setup with arbitrary undulator parameter, taking
into account a finite transverse size of the electron bunch.
Earlier studies considered a line density-distribution of
electrons instead. We focus our attention on the long-wavelength
asymptote (compared with resonance wavelength), at large distance
of the electron bunch from the undulator entrance compared to the
overtaking length, and for large vacuum-chamber size compared to
the typical transverse size of the field. These restrictions
define a parameter region of interest for practical applications.
We calculate a closed expression for impedance and wake function
that may be evaluated numerically in the most general case. Such
expression allows us to derive an analytical solution for a
Gaussian transverse and longitudinal bunch shape. Finally, we
study the feasibility of current-enhanced SASE schemes (ESASE)
recently proposed for LCLS, that fall well-within our
approximations. Numerical estimations presented in this paper
indicate that impedance-induced energy spread  is sufficient to
seriously degrade the FEL performance. Our conclusion is in
contrast with results in literature, where wake calculations for
the LCLS case are given in free-space, as if the presence of the
undulator were negligible.
\end{abstract}


\end{frontmatter}


\clearpage

\section{\label{sec:intro} Introduction}

Self-Amplified Spontaneous Emission Free Electron Lasers (SASE
FELs) \cite{ADD1, ADD2, ADD3} are nowadays considered as a unique
tool for production of intense, polarized, short-pulse radiation
tunable throughout the VUV and X-ray wavelength range, with peak
and average brilliance exceeding both modern Synchrotron Radiation
and Laser Plasma sources by many order of magnitudes \cite{ADD4,
ADD5, ADD6, ADD7, SLAC, XFEL}. Successful operation of SASE FELs
requires high quality (low emittance and low energy spread),
intense electron beams. One of the trends for SASE FELs is
production of ultra-short radiation pulses. These can be obtained
by exploiting electron bunches with an ultra-short charge
concentration (spike). At the FLASH\footnote{Free-Electron Laser
in Hamburg.} facility at DESY, Hamburg \cite{ADD4,ADD5,ADD6,ADD7},
electron bunches with sharp spikes have been produced in the
framework of a nonlinear bunch-compression scheme. Experimental
\cite{ADD4,ADD5,ADD6,ADD7} and theoretical \cite{ADD8, ADD9}
studies of FLASH operation have shown that properties of
ultra-short pulses are significantly influenced by collective
effects, the most important of them being space-charge effects.
Space-charge plays an important role in the beam-formation system,
in the drift space and also in a long undulator. Collective
effects might be crucially important for X-ray SASE FELs (XFELs)
as well.

This article presents a description of longitudinal wake fields in
XFELs. In particular, our study is of importance in connection
with novel schemes of radiation production, like Enhanced SASE
schemes (ESASE) \cite{ZHFA,CONF,ZHPE}. ESASE proposals rely on two
steps. First, the electron beam is modulated in energy by
interacting  with a GW-level optical laser in a modulator wiggler
placed in the accelerator section. Second, a dispersive section
transforms the energy modulation into density modulation,
eventually leading to a subfemtosecond-long spike in the beam
current before the entrance in the FEL undulator. The peak current
of this spike can reach tens of kA without emittance worsening,
because only a small charge is concentrated in the high-current
region. As the electron beam undergoes the SASE process, the
enhanced current part should saturate faster than the rest of the
bunch. Alternatively, the x-ray wavelength may be reduced, for a
fixed undulator length. Moreover ultra-short pulses (in the
attosecond range) are produced as a result of the presence of the
short lasing spike. Faster saturation of emission from the
enhanced-current spike also suggests that ESASE schemes may be
used to obtain saturation even in situations when beam parameters
deteriorate with respect to design values.

A detailed study of longitudinal wake fields arising after the
dispersive section, in particular dominant space-charge wake
fields is due in order to assess the magnitude of detrimental
effects on the FEL process. It is important to note that the
undulator parameter $K$ for XFEL setups obeys $K^2 \gg 1$. As a
result, the average longitudinal Lorentz factor $\bar{\gamma}_z =
\gamma/\sqrt{1+K^2/2}$ is such that $\bar{\gamma}_z^2 \ll
\gamma^2$, $\gamma$ being the Lorentz factor of the beam. Based on
$\bar{\gamma}_z^2 \ll \gamma^2$, we will demonstrate that the
presence of the undulator strongly influences the space-charge
wake. In contrast to this, in \cite{CONF,ZHPE}, wake calculations
for the LCLS case are given in free-space, as if the presence of
the undulator were negligible. Authors of references
\cite{CONF,ZHPE} incorrectly conclude that the FEL process is
basically unaffected by space-charge wakes.

This paper is devoted to the calculation of impedance and
longitudinal wake field in XFELs, with particular attention to the
LCLS case, for which ESASE schemes have been first proposed. This
means that we will restrict our attention to a very specific
region of parameters, discussed in the next Section
\ref{sec:disc}.  First, the longitudinal size of the beam is much
larger than the FEL wavelength. Second, electrons are assumed to
have travelled into the undulator for a distance longer than the
overtaking length. Third, effects of metallic surroundings can be
neglected. When the electron-beam size is larger than the
radiation diffraction size calculated from a single undulator
period\footnote{Of order $\sqrt{\lambdabar \lambda_w}$,
$\lambdabar$ being the reduced wavelength of coherent radiation
and $\lambdabar_w$ the undulator period.}, major simplifications
arise. In fact, radiation from the undulator is drastically
suppressed and calculations of impedance and wake function can be
performed considering a non-radiating beam, and thus accounting
for space-charge interactions only. Then, space-charge impedance
and wake function is found to reproduce the free-space case. Only,
the Lorentz factor $\gamma$ must be consistently substituted with
the average longitudinal Lorentz factor $\bar{\gamma}_z$. In
Section \ref{sec:main} we derive the electric field that will be
used to calculate impedance and wake. In Section \ref{sub:wake} we
introduce concepts of impedance and wake field. Fields are
calculated in Section \ref{sec:main}, while impedances and wakes
are respectively dealt with in Section \ref{sec:impe} and Section
\ref{sub:resu}. Then, in Section \ref{sec:expl}, we apply our
theory to the ESASE setup referring to the LCLS facility. We
calculate the energy chirp associated with wakes inside the
undulator and between dispersive section and undulator.
Subsequently, the magnitude of their effect is estimated by
calculating the linear energy chirp parameter \cite{KRIS,SALC}. We
find that the gain of the FEL process is sensibly reduced, and
that longitudinal wake fields constitute a reason of concern
regarding the practical realization of ESASE schemes.  Conclusions
end our treatment in Section \ref{sec:conc}.

\section{\label{sec:disc} Parameter-space of the problem}

As has been said in the Introduction, results of this paper can be
applied to calculate effects of wake fields in
planar\footnote{Although we presented final expressions of our
theory in the case of a planar undulator, there are no specific
effects related with the choice of a planar undulator. Our work
may be straightforwardly extended to the case of a helical
undulator as well.}  undulators under a specific choice of
parameters corresponding to an XFEL system. Quantities of interest
are defined once the bunch and the undulator system are specified.
The bunch is characterized by an rms length $\sigma_z$, a
transverse rms dimension $\sigma_\bot$ and Lorentz factor
$\gamma$. Moreover, we define the undulator period $\lambda_w$,
the vacuum chamber transverse dimension $a$ and the undulator
parameter $K$, where $K=\lambda_w e H_w/(2\pi m_e c^2)$, $(-e)$
being the negative electron charge, $H_w$ the peak undulator
magnetic field on-axis, and $m_e$ the rest mass of the electron.
Finally, $L_s$ is the saturation length of the FEL process.

The bunch length $\sigma_z$ corresponds to the reduced wavelength
of the coherent field generated by the bunch, $\sigma_z \simeq
\lambdabar \equiv \lambda/(2\pi)$. This (reduced) wavelength is
much longer than the reduced resonant wavelength $\lambdabar_r
\simeq \lambdabar_w/(2 \bar{\gamma}_z^2)$, where $\lambdabar_w
=\lambda_w/(2\pi)$, and $\bar{\gamma}_z = \gamma/\sqrt{1+K^2/2}$
is the already defined average longitudinal Lorentz factor. This
means $\lambdabar \gg \lambdabar_r$.

The overtaking length is defined by the quantity $2
\bar{\gamma}_z^2 \lambdabar$. When the bunch has travelled inside
the undulator for more than $2 \bar{\gamma}_z^2 \lambdabar$ a
steady state is reached, and asymptotic expressions for the wake
fields can be given. In the present study we will work with such
asymptotic expressions only. This means that the saturation length
of the FEL process,  $L_s$, must be much longer than the
overtaking length, i.e. $L_s \gg 2 \bar{\gamma}_z^2 \lambdabar$.

Also, in this paper we will neglect the presence of the vacuum
chamber. This is possible when the vacuum chamber dimension is
much larger than $\bar{\gamma}_z \lambdabar$, i.e. $a \gg
\bar{\gamma}_z \lambdabar$, a typical transverse dimension
associated with the coherent field, that is verified for ESASE
XFEL setups.

Summing up, we will work under the following constraints:

\begin{eqnarray}
&& \lambdabar ~~\gg ~\lambdabar_r ~~~~~~,\cr && L_s~ \gg~ 2
\bar{\gamma}_z^2 \lambdabar ~,\cr && a ~~~\gg~ \bar{\gamma_z}
\lambdabar ~~~. \label{cons1}
\end{eqnarray}
Based on conditions in (\ref{cons1}), we will develop a theory of
wake fields from undulators in XFELs. In particular, the first
assumption greatly simplifies our consideration allowing for a
long-wavelength asymptotic treatment. Under the second and the
third assumption we will be able to present an expression for the
impedance in terms of a double convolution involving the charge
density distribution and Bessel functions. Similarly, an
analytical expression for the wake function could be given.
However, it will not be necessary to explicitly calculate this
expression. In fact, when discussing practical applications, we
will work in the asymptotic case

\begin{eqnarray}
\sigma_\bot^2 \gg ~\lambdabar \lambdabar_w ~. \label{cons2}
\end{eqnarray}
Extra-condition (\ref{cons2}) greatly simplifies the treatment of
wake fields. Our results can be directly applied to realistic
situations as the ESASE scheme analyzed in Section \ref{sec:expl},
where we will refer, explicitly, to the LCLS case.

\section{\label{sec:main} Field calculation}

Calculation of longitudinal wake field and impedance from an FEL
undulator is subject to the characterization of the electric field
generated at a given position (that is the position of a test
electron) by the entire bunch.

We perform an analysis in terms of harmonics, i.e. we consider
sinusoidal dependence of the electric field of the kind $\vec{E} =
\vec{\bar{E}}(\vec{r},\omega) \exp[-i\omega t]+C.C.$, the symbol
"C.C." indicating complex conjugation\footnote{In the following,
for simplicity, we will consider $\omega > 0$. Expressions for the
field at negative values of $\omega$ can be obtained based on the
property $\vec{\bar{E}}(-\omega) = \vec{\bar{E}}^*(\omega)$
starting from explicit expressions for $\vec{\bar{E}}$ at
$\omega>0$.}. Here $t$ is the time, $\omega = 2\pi c/\lambda$ is
the frequency, with $c$ the speed of light in vacuum. The complex
amplitude $\vec{\bar{E}}(\vec{r},\omega)$ can actually be
considered as the representation of the electric field in the
space-frequency domain, and it will be referred to as "the field".

We assume that particles proceed along an undulator, under the
constraints discussed in Section \ref{sec:disc}.

\subsection{\label{sub:tran} Transverse field}

The transverse field $\vec{\bar{E}}_\bot$ can be treated in terms
of Paraxial Maxwell's  equations in the space-frequency domain
(see e.g. \cite{OURF,OURI}). From the paraxial approximation
follows that the electric field envelope $\vec{\widetilde{E}}_\bot
= \vec{\bar{E}}_\bot \exp{[-i\omega z/c]}$ does not vary much
along $z$ on the scale of the reduced wavelength
$\lambdabar=\lambda/(2\pi)$. As a result, the following field
equation holds:

\begin{eqnarray} \mathcal{D}
\left[\vec{\widetilde{E}}_\bot(z,\vec{r}_\bot,\omega)\right] =
\vec{g}(z, \vec{r}_\bot,\omega) ~,\label{field1}
\end{eqnarray}
where the differential operator $\mathcal{D}$ is defined by

\begin{eqnarray}
\mathcal{D} \equiv \left({\nabla_\bot}^2 + {2 i \omega \over{c}}
{\partial\over{\partial z}}\right) ~,\label{Oop}
\end{eqnarray}
${\nabla_\bot}^2$ being the Laplacian operator over transverse
cartesian coordinates. Eq. (\ref{field1}) is Maxwell's equation in
paraxial approximation. The source-term vector $\vec{g}(z,
\vec{r}_\bot)$ is specified by the trajectory of the source
electrons, and can be written in terms of the Fourier transform of
the transverse current density, $\vec{\bar{j}}_\bot
(z,\vec{r}_\bot,\omega)$, and of the charge density,
$\bar{\rho}(z,\vec{r}_\bot,\omega)$, as

\begin{eqnarray}
\vec{g} = && - {4 \pi} \exp\left[-\frac{i \omega z}{c}\right]
\left(\frac{i\omega}{c^2}\vec{\bar{j}}_\bot -\vec{\nabla}_\bot
\bar{\rho}\right)  ~. \label{fv}
\end{eqnarray}
$\vec{\bar{j}}_\bot$ and $\bar{\rho}$ are regarded as given data.
In this paper we will treat $\vec{\bar{j}}_\bot$ and $\bar{\rho}$
as macroscopic quantities, without investigating individual
electron contributions.  We consider transverse and longitudinal
distribution densities of the current constant through the
undulator.  In the time domain, we may write the charge density
$\rho(\vec{r},t)$ and the current density $\vec{j}(\vec{r},t)$ as

\begin{eqnarray}
\rho(\vec{r},t) =  \frac{1}{v_{oz}(z)}
\rho_\bot(\vec{r}_\bot-\vec{r'}_{o\bot}(z))
f\left(t-\frac{s_o(z)}{v_o}\right) \label{charge}
\end{eqnarray}
and

\begin{eqnarray}
\vec{j}(\vec{r},t) &=& \frac{1}{v_{oz}(z)} \vec{v}(z)
\rho_\bot(\vec{r}_\bot-\vec{r'}_{o\bot}(z))f\left(t-\frac{s_o(z)}{v_o}\right)
~. \cr&& \label{curr}
\end{eqnarray}
The quantity $\rho_\bot$ has the meaning of transverse electron
beam distribution, while $f$ is the longitudinal charge density
distribution.  $\vec{r'}_{o \bot}(z)$, $s_o(z)$ and $v_o$ pertain
a reference electron with Lorentz factor $\gamma$ that is injected
on axis with no deflection and is guided by the undulator field
only. Such electron follows a trajectory specified by
$\vec{r'}_{o\bot}(z)= r'_{ox} \vec{e}_x+r'_{oy} \vec{e}_y$,
$\vec{e}_x$ and $\vec{e}_y$ being the unit vectors in the
horizontal and vertical directions respectively, with

\begin{eqnarray}
&&r'_{ox}(z) = \frac{ K }{\gamma k_w} \cos(k_w z) = r_w \cos(k_w
z) ~~,~~~ r'_{oy}(z) = 0 ~, \label{rhel0}
\end{eqnarray}
where we defined the transverse amplitude of oscillations $r_w = K
/(\gamma k_w) $. The corresponding velocity is indicated with
$\vec{v}_{o\bot}(z)= v_{ox} \vec{e}_x+v_{oy} \vec{e}_y$:

\begin{eqnarray}
&&v_{ox}(z) = -  \frac{ K c}{\gamma} \sin(k_w z) ~~,~~~v_{oy}(z) =
0 ~. \label{vhel0}
\end{eqnarray}
Finally, $s_o(z)$ is the curvilinear abscissa measured along the
trajectory of the reference particle.

Note that, according to Eq. (\ref{charge}) and Eq. (\ref{curr}),
$\vec{{j}} = \vec{v}_{o} {\rho}$. In fact, for each particle in
the beam $\delta \gamma/\gamma \ll 1$. Therefore we can neglect
differences between the average transverse velocity of electrons
$\langle \vec{v} \rangle$ and $\vec{v}_{o}$.

In the space-frequency domain, Eq. (\ref{charge}) and Eq.
(\ref{curr}) transform to:

\begin{equation}
\bar{\rho}(\vec{r}_\bot,z,\omega) =
\rho_o\left(\vec{r}_\bot-\vec{r'}_{o\bot}(z)\right)
\bar{f}(\omega) \exp\left[{i \omega s_o(z)/v_o}\right]
\label{charge2tr}
\end{equation}
and

\begin{equation}
\vec{\bar{j}}(\vec{r}_\bot,z,\omega) =  \vec{v}_o(z)
\rho_o\left(\vec{r}_\bot-\vec{r'}_{o\bot}(z)\right)
\bar{f}(\omega)\exp\left[{i \omega s_o(z)/v_o}\right]~,
\label{curr2tr}
\end{equation}
where, for simplicity, we introduced the symbol

\begin{eqnarray}
\rho_o(\vec{r}_\bot) = \frac{1}{v_{oz}(z)}
\rho_\bot(\vec{r}_\bot)~. \label{rhoodef}
\end{eqnarray}
It should be remarked that $\bar{\rho}$ and $\vec{\bar{j}} =
\bar{\rho} \vec{v}_o$ satisfy the continuity equation. In other
words, one can find $\vec{\nabla} \cdot \vec{\bar{j}} = i\omega
\bar{\rho}$.

We note that for a generic motion one has

\begin{equation}
\omega \left({s_o(z_2)-s_o(z_1)\over{v}}-{z_2-z_1\over{c}}\right)
= \int_{z_1}^{z_2} d \bar{z} \frac{\omega}{2 \gamma_z^2(\bar{z})
c}~, \label{moregen}
\end{equation}
Also,

\begin{eqnarray}
\int_0^z \frac{\omega}{2c\gamma_z^2(\bar{z}) } d\bar{z}=
\frac{\omega}{2 c \bar{\gamma}_z^2}z - \frac{\omega K^2}{8
\gamma^2 k_w c} \sin(2 k_w z) \simeq \frac{\omega}{2 c
\bar{\gamma}_z^2}z~,\label{phasep}
\end{eqnarray}
where the average longitudinal Lorentz factor $\bar{\gamma}_z$ is
defined as

\begin{equation}
\bar{\gamma}_z = \frac{\gamma}{\sqrt{1+K^2/2}}~. \label{bargz}
\end{equation}
The approximate equality in Eq. (\ref{phasep}) follows from the
fact that we are interested in wavelengths $\omega \ll \omega_r$,
where the fundamental $\omega_r = 2 k_w c \bar{\gamma}_z^2$ is
fixed imposing resonance condition between electric field and
reference particle. The term in $\sin(2 k_w z')$ is of order
$\omega/\omega_r$, that is our accuracy, and can be neglected
everywhere.

With the help of Eq. (\ref{charge2tr}) and Eq. (\ref{curr2tr}),
Eq. (\ref{fv}) can be presented as (see also \cite{OURH}):

\begin{eqnarray}
\vec{g} = && - {4 \pi} \exp\left[i \int_{0}^{z} d \bar{z} \frac{
\omega }{2 \bar{\gamma}_z^2  c}\right]
\left[\frac{i\omega}{c^2}\vec{v}_{o\bot}(z) -\vec{\nabla}_\bot
\right]{\rho_o}\left(\vec{r}_\bot - \vec{r'}_{o\bot}(z)\right)
\bar{f}(\omega) ~.\cr && \label{fvtf}
\end{eqnarray}
We find an exact solution of Eq. (\ref{Oop}) without any other
assumption about the parameters of the problem. A Green's function
for Eq. (\ref{Oop}), namely the solution corresponding to the unit
point source can be written as (see \cite{OURF}):

\begin{eqnarray}
G(z-z';\vec{r_{\bot}}-\vec{r'_\bot}) &=& -{1\over{4\pi (z-z')}}
\exp\left\{ i\omega{\mid \vec{r_{\bot}}
-\vec{r'_\bot}\mid^2\over{2c (z-z')}}\right\}\label{green}~,
\end{eqnarray}
assuming $z-z' > 0$. When $z-z' < 0$ the paraxial approximation
does not hold, and the paraxial wave equation Eq. (\ref{field1})
should be substituted, in the space-frequency domain, by a more
general Helmholtz equation. Yet, the radiation formation length
for $z - z'<0$ is very short with respect to the case $z - z' >0$,
i.e. we can neglect contributions from sources located at $z-z'
<0$. 

Thus, after integration by parts, we obtain the solution

\begin{eqnarray}
\vec{\widetilde{E}}_{\bot}(z, \vec{r}_{\bot}) &=& \frac{i \omega
}{c} \int_{0}^{z} dz'   \int d \vec{r'}_{\bot} \exp\left\{i
\omega\left[\frac{\mid \vec{r}_{\bot }-\vec{r'}_\bot \mid^2}{2c
(z-z')}\right]+  i \int_{0}^{z'} d \bar{z}\frac{ \omega }{2 c
\gamma_z^2(\bar{z})} \right\} \cr&&\times \frac{1}{z-z'}
{\rho_o}\left(\vec{r'}_\bot-\vec{r'}_{\bot o}(z')\right)
\bar{f}(\omega) \left(\frac{\vec{v}_{\bot}(z')}{c}
-\frac{\vec{r}_{\bot}-\vec{r'}_{\bot}}{z-z'}\right)~. \cr &&
\label{generalfin}
\end{eqnarray}
Eq. (\ref{generalfin}) describes the field at any position $z$.
Note that ${\rho_o}$ depends on the difference $\vec{r'}_\bot
-\vec{r'}_{\bot o}(z')$. This dependence is important concerning
the effect studied in this paper, as it will be seen later on.

Eq. (\ref{generalfin}) consists of two terms: one in
$\vec{v}_\bot$ and the other in $\vec{r}_{\bot} - \vec{r'}_\bot$.
We will sometimes name the first term the "current term"
$\vec{\widetilde{E}}_{\bot c}$, while the second will be indicated
as the gradient term $\vec{\widetilde{E}}_{\bot g}$. With the help
of Eq. (\ref{rhel0}), Eq. (\ref{vhel0}) and Eq. (\ref{phasep}) we
can re-write Eq. (\ref{generalfin}) as

\begin{eqnarray}
&&\vec{\widetilde{E}}_{\bot}(z, \vec{r}_{\bot}) = -\frac{i \omega
}{c} \int_{0}^{z} dz' \frac{1}{z-z'}  \int d \vec{r'}_{\bot}
\left(\frac{K \vec{e}_x}{\gamma}\sin(k_w z')
+\frac{\vec{r}_{\bot}-\vec{r'}_{\bot}}{z-z'}\right)\cr&&\times
{\rho_o}\left(\vec{r'}_\bot-r_w \cos(k_w z') \vec{e}_x\right)
\bar{f}(\omega) \exp\left\{i \omega\left[\frac{\mid
\vec{r}_{\bot}-\vec{r'}_\bot \mid^2}{2c (z-z')}\right]+
\frac{i\omega z'}{2 c \bar{\gamma}_z^2} \right\} ~. \cr &&
\label{generalfin2}
\end{eqnarray}

\subsection{\label{longu} Longitudinal field}

A similar expression can be found for the longitudinal field.
Since $v_z(z) \simeq c$, we can write the longitudinal equivalent
of Eq. (\ref{fv}) as

\begin{eqnarray}
{g_z} = && - {4 \pi} \exp\left[-\frac{i \omega z}{c}\right]
\left(\frac{i\omega}{c^2} {\bar{j}}_z -\partial_z
\bar{\rho}\right) ~, \label{fv2}
\end{eqnarray}
that is

\begin{eqnarray}
g_z = && - {4 \pi} \exp\left[i \int_{0}^{z} d \bar{z} \frac{
\omega }{2 \bar{\gamma}_z^2  c}\right]
\left[-\frac{i\omega}{\bar{\gamma}_z^2 c}
-\frac{\partial}{\partial z} \right]{\rho_o}\left(\vec{r}_\bot -
\vec{r'}_{o\bot}(z)\right) \bar{f}(\omega) ~,\cr && \label{fztf}
\end{eqnarray}
having used the fact that $v_{oz}(z) \simeq c$.

It follows that the longitudinal component of the field, can be
written analogously to Eq. (\ref{generalfin}) as

\begin{eqnarray}
{\widetilde{E}}_{z}(z, \vec{r}_{\bot}) &=&  \int_{0}^{z} dz' \int
d \vec{r'}_{\bot} \exp\left\{i \omega\left[\frac{\mid
\vec{r}_{\bot}-\vec{r'}_\bot \mid^2}{2c (z-z')}\right]+  i
\int_{0}^{z'} d \bar{z}\frac{ \omega }{2 c \gamma_z^2(\bar{z})}
\right\} \cr && \times \frac{1}{z-z'} \left[-\frac{i \omega
}{\bar{\gamma}_z^2 c} -\frac{\partial}{\partial z'}\right]
{\rho_o}\left(\vec{r'}_\bot-\vec{r'}_{\bot o}(z')\right)
\bar{f}(\omega)~. \cr && \label{generalfinz}
\end{eqnarray}
Note that the integral in Eq. (\ref{generalfinz}) is performed for
$z'$ ranging from $0$ up to $z$ exactly as the integral in Eq.
(\ref{generalfin2}), for the same reasons.

Use of Eq. (\ref{rhel0}), Eq. (\ref{vhel0}) and Eq. (\ref{phasep})
allow to re-write Eq. (\ref{generalfinz}) as

\begin{eqnarray}
&&{\widetilde{E}}_{z}(z, \vec{r}_{\bot}) =  \int_{0}^{z} dz'
\frac{1}{z-z'}  \int d \vec{r'}_{\bot} \exp\left\{i
\omega\left[\frac{\mid \vec{r}_{\bot}-\vec{r'}_\bot \mid^2}{2c
(z-z')}\right]+ \frac{i\omega z'}{2 c \bar{\gamma}_z^2}
\right\}\cr&&\times \left[-\frac{i \omega }{\bar{\gamma}_z^2
c}-\frac{K}{\gamma}\sin(k_w z')\frac{\partial}{\partial
[x'-r'_{ox}(z')] }\right]{\rho_o}\left(\vec{r'}_\bot-r_w \cos(k_w
z') \vec{e}_x\right) \bar{f}(\omega) ~. \cr &&
\label{generalfinz2}
\end{eqnarray}
Finally, integration by parts of the term in $\partial/\{\partial
[x'-r'_{ox}(z')]\}$ gives

\begin{eqnarray}
&&{\widetilde{E}}_{z}(z, \vec{r}_{\bot}) = - \frac{i \omega }{c}
\int_{0}^{z} dz' \frac{1}{z-z'}  \int d \vec{r'}_{\bot}
\left[\frac{1}{\bar{\gamma}_z^2}+ \frac{K}{\gamma}\sin(k_w
z')\frac{x-x'}{z-z' }\right]\cr&&\times
{\rho_o}\left(\vec{r'}_\bot-r_w \cos(k_w z') \vec{e}_x\right)
\bar{f}(\omega) \exp\left\{i \omega\left[\frac{\mid
\vec{r}_{\bot}-\vec{r'}_\bot \mid^2}{2c (z-z')}\right]+
\frac{i\omega z'}{2 c \bar{\gamma}_z^2} \right\} ~.
\label{generalfinz2b}
\end{eqnarray}
Note that, in contrast with Eq. (\ref{generalfin2}), we cannot
clearly distinguish between "current" and "gradient" terms in Eq.
(\ref{generalfinz2b}): the term in $\sin(k_w z')$ can be traced
back to the gradient of the charge density, but the one in
$1/\bar{\gamma}_z^2$ is a combination between current and gradient
term.

\subsection{\label{sub:role} Perturbation theory}

It is possible to analyze Eq. (\ref{generalfin2}) and Eq.
(\ref{generalfinz2b}) in the framework of a perturbation theory,
based on expansion in the small parameter $\lambdabar_r/\lambdabar
\ll 1$ according to the first of conditions (\ref{cons1}). This
allows simplified treatment of impedance and wakes.

The first step towards this direction is a presentation of of
$\vec{\widetilde{E}}_\bot$ and $\widetilde{E}_z$ with the help of
the following expansions in plane waves:

\begin{eqnarray}
&&\frac{1}{\left(z-z'\right)}\exp\left\{i \omega\left[\frac{\mid
\vec{r}_{\bot}-\vec{r'}_\bot \mid^2}{2c (z-z')}\right]\right\} =
\cr &&\frac{i c}{2\pi \omega} \int d\vec{k}_\bot \exp\left[-i
\vec{k}_\bot \cdot \left( \vec{r}_\bot -\vec{r'}_\bot
\right)\right] \exp\left[\frac{i k_\bot^2 c
}{2\omega}(z'-z)\right]\label{expand1}
\end{eqnarray}
and

\begin{eqnarray}
&& \frac{\vec{r}_{\bot}-\vec{r'}_\bot}{(z-z')^2}\exp\left\{i
\omega\left[\frac{\mid \vec{r}_{\bot}-\vec{r'}_\bot \mid^2}{2c
(z-z')}\right]\right\} = - \frac{i c^2}{2 \pi \omega^2}\cr && \int
d\vec{k}_\bot \vec{k}_\bot \exp\left[-i \vec{k}_\bot \cdot \left(
\vec{r}_\bot -\vec{r'}_\bot \right)\right] \exp\left[\frac{i
k_\bot^2 c }{2\omega}(z'-z)\right]~.\label{expand2}
\end{eqnarray}
We obtain

\begin{eqnarray}
&&\vec{\widetilde{E}}_{\bot}(z, \vec{r}_{\bot}) =
\frac{\bar{f}(\omega)}{2\pi} \int d\vec{k}_\bot\int_{0}^{z} dz'
\int d \vec{r'}_{\bot} \left[\frac{K \vec{e}_x}{\gamma}\sin(k_w
z') -\frac{c \vec{k}_\bot}{\omega}\right] \exp\left[\frac{i\omega
z'}{2 c \bar{\gamma}_z^2} \right]\cr&&\times
{\rho_o}\left(\vec{r'}_\bot-r_w \cos(k_w z') \vec{e}_x\right)
\exp\left[-i \vec{k}_\bot \cdot \left( \vec{r}_\bot -\vec{r'}_\bot
\right)\right] \exp\left[\frac{i k_\bot^2 c
}{2\omega}(z'-z)\right]  \label{perpexp}
\end{eqnarray}
and
\begin{eqnarray}
&&{\widetilde{E}}_{z}(z, \vec{r}_{\bot}) =
\frac{\bar{f}(\omega)}{2\pi} \int d\vec{k}_\bot \int_{0}^{z} dz'
\int d \vec{r'}_{\bot} \left[\frac{1}{\bar{\gamma}_z^2}- \frac{K c
k_x}{\gamma\omega }\sin(k_w z')\right]\exp\left[ \frac{i\omega
z'}{2 c \bar{\gamma}_z^2} \right]\cr&&\times
{\rho_o}\left(\vec{r'}_\bot-r_w \cos(k_w z') \vec{e}_x\right)
\exp\left[-i \vec{k}_\bot \cdot \left( \vec{r}_\bot -\vec{r'}_\bot
\right)\right] \exp\left[\frac{i k_\bot^2 c
}{2\omega}(z'-z)\right]~. \label{longexp}
\end{eqnarray}
Performing a change of variables $\vec{r'}_\bot\longrightarrow
\vec{r'}_\bot-r_w \cos(k_w z') \vec{e}_x$ and introducing notation
$\vec{\theta} = \vec{k}_\bot c/\omega$ we re-write Eq.
(\ref{perpexp}) and Eq. (\ref{longexp}) as

\begin{eqnarray}
&&\vec{\widetilde{E}}_{\bot}(z, \vec{r}_{\bot}) = \frac{\omega^2
\bar{f}(\omega)}{2\pi c^2} \int d\vec{\theta}\int_{0}^{z} dz' \int
d \vec{r'}_{\bot} \left[\frac{K \vec{e}_x}{\gamma}\sin(k_w z') -
\vec{\theta}\right] \exp\left[\frac{i\omega z'}{2 c
\bar{\gamma}_z^2} \right]\cr&&\times
{\rho_o}\left(\vec{r'}_\bot\right) \exp\left[-\frac{i \omega}{c}
\vec{\theta} \cdot \left( \vec{r}_\bot -\vec{r'}_\bot-r_w \cos(k_w
z') \vec{e}_x \right)\right] \exp\left[\frac{i \omega \theta^2}{2
c }(z'-z)\right]   \label{perpexp2}
\end{eqnarray}
and
\begin{eqnarray}
&&{\widetilde{E}}_{z}(z, \vec{r}_{\bot}) = \frac{\omega^2
\bar{f}(\omega)}{2\pi c^2} \int d\vec{\theta} \int_{0}^{z} dz'
\int d \vec{r'}_{\bot} \left[\frac{1}{\bar{\gamma}_z^2} - \frac{K
\theta_x}{\gamma}\sin(k_w z')\right]\exp\left[ \frac{i\omega z'}{2
c \bar{\gamma}_z^2} \right]\cr&&\times
{\rho_o}\left(\vec{r'}_\bot\right) \exp\left[-\frac{i \omega}{c}
\vec{\theta} \cdot \left( \vec{r}_\bot -\vec{r'}_\bot-r_w \cos(k_w
z') \vec{e}_x \right)\right] \exp\left[\frac{i \omega \theta^2 }{2
c }(z'-z)\right]~. \label{longexp2}
\end{eqnarray}
%
Note that the maximal range of angles $\theta_{x,y}$ is limited by
the last exponential function in Eq. (\ref{perpexp2}) and Eq.
(\ref{longexp2}) and by the fact that $z-z' \gtrsim \lambdabar_w$.
It follows that $\theta_{x,y}$ cannot be larger than about
$\sqrt{\lambdabar/\lambdabar_w}$.  Then, the trigonometric terms
inside the exponential functions in both Eq. (\ref{perpexp2}) and
Eq. (\ref{longexp2}) are of magnitude $\omega \theta_x r_w/c
\lesssim \sqrt{\lambdabar_r/\lambdabar} \ll 1$. It follows that we
may expand $\exp\{i \omega r_w \theta_x \cos[k_w z']/c\} \simeq
1+i \omega r_w \theta_x  \cos[k_w z']/c$. Using exponential
representation for all trigonometric functions we obtain:

\begin{eqnarray}
&&\vec{\widetilde{E}}_{\bot}(z, \vec{r}_{\bot}) = \frac{\omega^2
\bar{f}(\omega)}{2\pi c^2} \int d\vec{\theta} \exp\left[-\frac{i
\omega \theta^2 z}{2 c }\right] \int_{0}^{z} dz' \int d
\vec{r'}_{\bot} \cr && \times{\rho_o}\left(\vec{r'}_\bot\right)
\exp\left[-\frac{i \omega}{c} \vec{\theta} \cdot \left(
\vec{r}_\bot -\vec{r'}_\bot\right)\right]  \left\{\frac{K
\vec{e}_x}{2 i \gamma}\left[\exp(i k_w z')-\exp(-ik_w z')\right] -
\vec{\theta}\right\}\cr &&\times \left\{1+\frac{i \omega \theta_x
r_w }{2 c} \left[\exp(i k_w z')+\exp(-i k_w z')\right]\right\}
\exp\left[\frac{i \omega \theta^2 z'}{2 c }\right]\exp\left[
\frac{i\omega z'}{2 c \bar{\gamma}_z^2} \right]~. \label{perpexp3}
\end{eqnarray}
and

\begin{eqnarray}
&&{\widetilde{E}}_{z}(z, \vec{r}_{\bot}) = \frac{\omega^2
\bar{f}(\omega)}{2\pi c^2} \int d\vec{\theta}  \exp\left[-\frac{i
\omega \theta^2 z}{2 c }\right] \int_{0}^{z} dz' \int d
\vec{r'}_{\bot} \cr&&\times
{\rho_o}\left(\vec{r'}_\bot\right)\exp\left[-\frac{i \omega}{c}
\vec{\theta} \cdot \left( \vec{r}_\bot -\vec{r'}_\bot
\right)\right] \left\{\frac{1}{\bar{\gamma}_z^2}- \frac{K
\theta_x}{2 i \gamma}\left[\exp(i k_w z')-\exp(-ik_w
z')\right]\right\}\cr &&\times \left\{1+\frac{i \omega \theta_x
r_w }{2 c} \left[\exp(i k_w z')+\exp(-i k_w
z')\right]\right\}\exp\left[\frac{i \omega \theta^2 z'}{2 c
}\right]\exp\left[ \frac{i\omega z'}{2 c \bar{\gamma}_z^2}
\right]~. \label{longexp3}
\end{eqnarray}
Eq. (\ref{perpexp3}) and Eq. (\ref{longexp3}) have been found
exploiting the small parameter $\lambdabar_r/\lambdabar$. In both
Eq. (\ref{perpexp3}) and Eq. (\ref{longexp3}) products of factors
within $\{...\}$ brackets are of the form $\exp[\pm i p k_w z']$,
with $p=0,1,2...$, terms for $p > 1$ being obtainable considering
higher orders in $\sqrt{\lambdabar/\lambdabar_w}$ in the previous
expansion of the exponential of trigonometric function. Note that
when $p=0$, the magnitude of $\theta_{x,y}$ can be estimated from
the last two exponential functions in Eq. (\ref{perpexp3}) and Eq.
(\ref{longexp3}), giving characteristic a scale $\theta_{x,y} \sim
1/\bar{\gamma}_z$. For other values of $p$, instead, we have a
characteristic scale $\theta_{x,y} \sim
\sqrt{\lambdabar/\lambdabar_w}$.

Consider first Eq. (\ref{perpexp3}). Magnitudes of factors within
$\{...\}$ brackets are $K/(2 \gamma)$ and $\theta$ for the first
bracket, $1$ and $\omega \theta_x r_w/(2c)$ for the second
bracket. Terms of the form $\exp[\pm i p k_w z']$ with $p=0$ can
have magnitudes $\theta \sim 1/\bar{\gamma}_z$ or $[K
/(2\gamma)]\cdot [\omega \theta_x r_w/(2c)]\sim K^2 \lambdabar_r
\bar{\gamma}_z/(\gamma^2 \lambdabar)$, this last kind being
negligible. When $p=1$ terms have magnitudes $K/(2\gamma)$ or
$\theta \omega \theta_x r_w/(2c) \sim K/\gamma$, and both kinds
have to be kept. Similarly, it can be shown that all other values
of $p$ give negligible terms.

Consider now Eq. (\ref{longexp3}). Magnitudes of factors within
$\{...\}$ brackets are $1/\bar{\gamma}_z^2$ and $K\theta_x/(2
\gamma)$ for the first bracket, $1$ and $\omega \theta_x r_w/(2c)$
for the second bracket. Terms of the form $\exp[\pm i p k_w z']$
with $p=0$ can have magnitudes $1/\bar{\gamma}_z^2$ or $[K
\theta_x/(2\gamma)]\cdot [\omega \theta_x r_w/(2c)]\sim K^2
\lambdabar_r /(\gamma^2 \lambdabar)$, this last kind being
negligible. When $p=1$ terms have magnitudes $K \theta_x/2\gamma
\sim K/\gamma \cdot \sqrt{\lambdabar/\lambdabar_w} $ or $\omega
\theta_x r_w/(2c \bar{\gamma}_z^2 ) \sim
\sqrt{\lambdabar_r/\lambdabar}\cdot1/(\bar{\gamma}_z\gamma)$, and
this last kind can be neglected. Similarly, it can be shown that
all other values of $p$ give negligible terms.

Altogether, we obtain the following expressions for transverse and
longitudinal field:

\begin{eqnarray}
&&\vec{\widetilde{E}}_{\bot}(z, \vec{r}_{\bot}) = \frac{\omega^2
\bar{f}(\omega)}{2\pi c^2} \int d\vec{\theta} \int_{0}^{z} dz'
\int d \vec{r'}_{\bot} \cr &&
\times{\rho_o}\left(\vec{r'}_\bot\right) \exp\left[-\frac{i
\omega}{c} \vec{\theta} \cdot \left( \vec{r}_\bot
-\vec{r'}_\bot\right)\right] \exp\left[\frac{i \omega \theta^2
(z'-z)}{2 c }\right]\exp\left[ \frac{i\omega z'}{2 c
\bar{\gamma}_z^2} \right] \cr && \times \left\{-\vec{\theta}
+\left[\frac{K \vec{e}_x}{2i\gamma}-\frac{i \omega \theta_x
r_w\vec{\theta}}{2c}\right]\exp(i k_w z')- \left[\frac{K
\vec{e}_x}{2i\gamma}+\frac{i \omega \theta_x
r_w\vec{\theta}}{2c}\right]\exp(-i k_w z')\right\}\cr &&
\label{perpexp4}
\end{eqnarray}
and

\begin{eqnarray}
&&{\widetilde{E}}_{z}(z, \vec{r}_{\bot}) = \frac{\omega^2
\bar{f}(\omega)}{2\pi c^2}   \int d\vec{\theta}  \int_{0}^{z} dz'
\int d \vec{r'}_{\bot}
{\rho_o}\left(\vec{r'}_\bot\right)\exp\left[-\frac{i \omega}{c}
\vec{\theta} \cdot \left( \vec{r}_\bot -\vec{r'}_\bot
\right)\right]\cr&&\times \exp\left[\frac{i \omega \theta^2
(z'-z)}{2 c }\right]\exp\left[ \frac{i\omega z'}{2 c
\bar{\gamma}_z^2} \right]
\left\{\frac{1}{\bar{\gamma}_z^2}-\frac{K \theta_x}{2 \gamma
i}\left[\exp(i k_w z')-\exp(-ik_w z')\right]\right\}~.\cr &&
\label{longexp4}
\end{eqnarray}
Eq. (\ref{perpexp4}) and Eq. (\ref{longexp4}) are the first order
result of our perturbation theory, where the small parameter
$\lambdabar_r/\lambdabar$ has been exploited through the expansion
of exponential functions in Eq. (\ref{perpexp2}) and Eq.
(\ref{longexp2}), and non-negligible terms are kept.

We now go back to the space-frequency domain performing the
integral in $d\vec{\theta}$ with the help of Eq. (\ref{expand1}),
Eq. (\ref{expand2}) and using also

\begin{eqnarray}
&&\int d\vec{\theta} \exp\left[-\frac{i \omega}{c} \vec{\theta}
\cdot \left( \vec{r}_\bot -\vec{r'}_\bot \right)\right]
\exp\left[\frac{i \omega \theta^2 (z'-z)}{2 c }\right] \theta_x^2
\cr && = \frac{2 c \pi \left[c(z-z')+i \omega(x-x')^2
\right]}{(z'-z)^3} \exp\left[\frac{i \omega \left|\vec{r}_\bot -
\vec{r'}_\bot \right|^2}{2c (z-z') }\right]~.\label{lastback}
\end{eqnarray}
Finally, we obtain:

\begin{eqnarray}
&&\vec{\widetilde{E}}_{\bot}(z, \vec{r}_{\bot}) = -\frac{i\omega
\bar{f}(\omega)}{c} \int d \vec{r'}_{\bot}
{\rho_o}\left(\vec{r'}_\bot\right) \exp\left[ \frac{i\omega z}{2 c
\bar{\gamma}_z^2}\right]\cr && \times \Bigg\{\exp[+i k_w z]
\int_{0}^{z} \frac{ dz'}{z-z'} \exp\left[i \omega \frac{\mid
\vec{r}_{\bot}-\vec{r'}_\bot \mid^2}{2c (z-z')} \right]
\left[+\frac{K \vec{e}_x}{2i\gamma}\exp[ik_w (z'-z)]\right] \cr
&&~~+ \exp[-i k_w z] \int_{0}^{z} \frac{ dz'}{z-z'} \exp\left[i
\omega \frac{\mid \vec{r}_{\bot}-\vec{r'}_\bot \mid^2}{2c (z-z')}
\right] \left[-\frac{K \vec{e}_x}{2i\gamma}\exp[ik_w(z- z')]
\right] \cr && ~~+\exp[+i k_w z] \int_{0}^{z} \frac{ dz'}{z-z'}
\exp\left[i \omega \frac{\mid \vec{r}_{\bot}-\vec{r'}_\bot
\mid^2}{2c (z-z')} \right] \cr &&
~~~~~~~~~~~~~~~~~~~~~~\times\left[ -\frac{r_w \vec{e}_x}{2 (z-z')}
- \frac{i\omega r_w(x-x')(\vec{r}_\bot-\vec{r'}_\bot)}{2
c(z-z')^2} \right]\exp[ik_w (z'-z)] \cr &&~~+ \exp[-i k_w z]
\int_{0}^{z} \frac{ dz'}{z-z'} \exp\left[i \omega \frac{\mid
\vec{r}_{\bot}-\vec{r'}_\bot \mid^2}{2c (z-z')} \right]\cr &&
~~~~~~~~~~~~~~~~~~~~~~\times\left[ -\frac{r_w \vec{e}_x}{2 (z-z')}
- \frac{i\omega r_w(x-x')(\vec{r}_\bot-\vec{r'}_\bot)}{2
c(z-z')^2} \right]\exp[ik_w (z-z')]  \cr && +\int_{0}^{z} \frac{
dz'}{z-z'} \exp\left[i \omega \frac{\mid
\vec{r}_{\bot}-\vec{r'}_\bot \mid^2}{2c (z-z')} \right] \exp\left[
\frac{i\omega (z'-z)}{2 c
\bar{\gamma}_z^2}\right]\left[\frac{\vec{r}_{\bot}-\vec{r'}_{\bot}}{z-z'}\right]
\Bigg\} \label{effp}
\end{eqnarray}
for the transverse field, and

\begin{eqnarray}
&&{\widetilde{E}}_{z}(z, \vec{r}_{\bot}) = - \frac{i \omega
\bar{f}(\omega)}{c} \int d \vec{r'}_{\bot}
{\rho_o}\left(\vec{r'}_\bot\right) \exp\left[\frac{i\omega z}{2 c
\bar{\gamma}_z^2}\right]\cr && \times \Bigg\{\exp[+i k_w
z]\int_{0}^{z} \frac{dz'}{z-z'} \exp\left[i \omega\frac{\mid
\vec{r}_{\bot}-\vec{r'}_\bot \mid^2}{2c (z-z')}\right]\left[
+\frac{K}{2 i \gamma}\frac{x-x'}{z-z'}\exp[i k_w (z'-z)]\right]
\cr && + \exp[-i k_w z] \int_{0}^{z} \frac{dz'}{z-z'} \exp\left[i
\omega\frac{\mid \vec{r}_{\bot}-\vec{r'}_\bot \mid^2}{2c
(z-z')}\right]\left[ -\frac{K}{2 i \gamma}\frac{x-x'}{z-z'
}\exp[-i k_w (z'-z)]\right]\cr && +\int_{0}^{z} \frac{dz'}{z-z'}
\exp\left[i \omega\frac{\mid \vec{r}_{\bot}-\vec{r'}_\bot
\mid^2}{2c (z-z')}\right]\exp\left[ \frac{i\omega (z'-z)}{2 c
\bar{\gamma}_z^2} \right]\frac{1}{\bar{\gamma}_z^2}\Bigg\}~
\label{effz}
\end{eqnarray}
for the longitudinal field. Here we neglected factors $\exp[i
\omega (z'-z)/(2 c \bar{\gamma}_z^2)]$ in integral terms in $dz'$
including $\exp[\pm k_w z]$, because $\omega \lambda_w / (2 c
\bar{\gamma}_z^2) \ll 1$.

Note that there exists a mathematical shortcut to obtain Eq.
(\ref{effp}) and Eq. (\ref{effz}) from Eq. (\ref{generalfin2}) and
Eq. (\ref{generalfinz2b}). In fact, if we perform a change of
variables $\vec{r'}_\bot\longrightarrow \vec{r'}_\bot-r_w \cos(k_w
z') \vec{e}_x$, we formally expand the Green's function
exponential $\exp\{i \omega [{\mid \vec{r}_{\bot}-\vec{r'}_\bot
-r_w \cos(k_w z') \vec{e}_x\mid^2}/{2c (z-z')}]\}$ to the first
order in $r_w$ and keep non-negligible first-harmonic terms in
$\exp[\pm i k_w z']$, we obtain  Eq. (\ref{effp}) and Eq.
(\ref{effz}).  We will regard it as a mnemonic rule, that will be
useful later on.

Fields are calculated under conditions (\ref{cons1}). In the limit
for $z \gg \bar{\gamma}_z^2 \lambdabar$, as we will see in the
next Section \ref{sub:expl}, integrals in $dz'$ depend on $z$ only
through phase factors, i.e. a steady state solution is reached.

Analysis of Eq. (\ref{effp}) and Eq. (\ref{effz}) presents an
interesting picture of the fields generated by the electron beam.
Eq. (\ref{effp}) and Eq. (\ref{effz}) consist of the sum of
integrals in $dz'$. Some include exponential factors $\exp[\pm k_w
z]$, other not.

Terms not including $\exp[\pm k_w z]$ (the last integrals in $dz'$
in both Eq. (\ref{effp}) and Eq. (\ref{effz})) oscillate, as a
function of $z$, on a scale $\lambdabar \bar{\gamma}_z^2$. The
field ${\vec{\bar{E}}}$ is given by $\vec{\bar{E}} \exp[i\omega
z/c]$. It follows that the phase velocity of terms not including
$\exp[\pm k_w z]$ is the same as that of the electron beam
harmonic $\bar{\rho}$. We can interpret this fact by saying that
this part of the field is entangled with the electron beam. It is
natural to identify these terms as space-charge terms. The
formation length of the space-charge field is determined by the
factor $\exp[i \omega (z'-z)/(2 c \bar{\gamma}_z^2)]$ under
integral sign, and amounts to $2 \lambdabar \bar{\gamma}_z^2$.
Similarly, the diffraction size of the space-charge field is given
by $\bar{\gamma}_z \lambdabar$.

Terms including $\exp[\pm i k_w z]$ are indicative of fields
${\vec{\bar{E}}}$ performing a cycle of oscillation on the scale
of an undulator period with respect to the electron-beam harmonic
$\bar{\rho}$. Phase velocity of terms including $\exp[+ i k_w z]$
is slower than that of the beam harmonic. These field terms have a
phase velocity slower than the speed of light. Phase velocity of
terms including $\exp[- i k_w z]$ is faster than that of the beam
harmonic. These field terms have a phase velocity faster than the
speed of light. We can interpret these facts by saying that these
parts of the field are not entangled with the electron beam. It is
natural to identify these terms as radiation terms. The formation
length of radiation field terms is determined by the factor
$\exp[\pm i k_w (z'-z)]$ under integral sign, and amounts to
$\lambdabar_w$. Similarly, the diffraction size of the
space-charge field is given by $\sqrt{\lambdabar \lambdabar_w}$.

It is interesting to trace each term in Eq. (\ref{effp}) and Eq.
(\ref{effz}) back to the source terms that originated them,
distinguishing between gradient and current terms. Considering Eq.
(\ref{effp}) it can be seen that the first and the second integral
are (radiative) current density terms. The third and the fourth
term are (radiative) gradient terms, while the last term is a
(space charge) gradient term. However, one can see from Eq.
(\ref{effz}) that the third integral is a (space-charge) term
originated from a mixture of gradient and current sources. Thus,
although the first and the second integral are (radiative)
gradient terms, it does not make sense to separately talk about
gradient and current term for the longitudinal component of the
field.

An interesting picture emerges, where radiation field and
space-charge field are treated on equal foot, through paraxial
Maxwell's equation. On the one hand, as we have seen, these fields
have different formation lengths, and different diffraction sizes.
On the other hand, our theory allows for generic transverse sizes
of the electron beam $\sigma_\bot$, that makes it possible to
compare $\sigma_\bot$ with both diffraction sizes, thus obtaining
different regimes. As we will see, when $\sigma_\bot \gg
\sqrt{\lambdabar \lambdabar_w}$, impedance and wakes are
essentially dominated by the longitudinal space-charge term. It is
important to remark, for future use in the next Sections, that
$\bar{\gamma}_z$ enters the expression of the space-charge field,
and not $\gamma$.

\subsection{\label{sub:expl} Explicit expressions for the field}

We can consider Eq. (\ref{effp}) and Eq. (\ref{effz}) as starting
point for our investigations, and calculate explicit expressions
for the field to be used later on in the calculation of the
impedance. First, we make a change in the integration variable
from $z'$ to $\xi \equiv z-z'$. In the limit for $z
\longrightarrow \infty$, corresponding to the second of conditions
(\ref{cons1}), i.e. $z \gg \bar{\gamma}_z^2 \lambdabar$, we can
write

\begin{eqnarray}
\vec{\widetilde{E}}_{\bot}(z, \vec{r}_{\bot}) &=& -\frac{i \omega
\bar{f}(\omega)}{c} \int d \vec{r'}_{\bot}
{\rho_o}\left(\vec{r'}_\bot\right) \exp\left[ \frac{i\omega z}{2 c
\bar{\gamma}_z^2}\right] \cr && \times\Bigg\{+\frac{K
\vec{e}_x}{2i\gamma} \exp[+i k_w z] \int_{0}^{\infty} \frac{ d
\xi}{\xi}  \exp\left[+i \omega \frac{\mid
\vec{r}_{\bot}-\vec{r'}_\bot \mid^2}{2c \xi}-i k_w \xi \right] \cr
&&~~~~~-\frac{K \vec{e}_x}{2i\gamma}  \exp[-i k_w z]
\int_{0}^{\infty} \frac{ d\xi}{\xi} \exp\left[+i \omega \frac{\mid
\vec{r}_{\bot}-\vec{r'}_\bot \mid^2}{2c \xi} + i k_w \xi \right]
\cr&& ~~~~~+\exp[+i k_w z] \Bigg[ +\frac{i c r_w \vec{e}_x}{2
\omega \mid \vec{r}_{\bot}-\vec{r'}_\bot \mid}\cdot
\frac{d}{d\left[\mid \vec{r}_{\bot}-\vec{r'}_\bot \mid\right]}\cr
&&~~~~~ + \frac{2 i c r_w}{\omega
}(x-x')(\vec{r}_\bot-\vec{r'}_\bot)\cdot \frac{d^2}{d \left[\mid
\vec{r}_{\bot}-\vec{r'}_\bot \mid^2\right]^2} \Bigg]\cr
&&~~~~~~~~~~~~~~~~~~~~~~~~~~~~~~~\times \int_{0}^{\infty} \frac{
d\xi}{\xi} \exp\left[+i \omega \frac{\mid
\vec{r}_{\bot}-\vec{r'}_\bot \mid^2}{2c \xi}-ik_w \xi \right] \cr
&&~~~~~+ \exp[-i k_w z] \Bigg[ +\frac{i c r_w \vec{e}_x}{2 \omega
\mid \vec{r}_{\bot}-\vec{r'}_\bot \mid}\cdot \frac{d}{d\left[\mid
\vec{r}_{\bot}-\vec{r'}_\bot \mid\right]}\cr &&~~~~~ + \frac{2 i c
r_w}{ \omega }(x-x')(\vec{r}_\bot-\vec{r'}_\bot)\cdot \frac{d^2}{d
\left[\mid \vec{r}_{\bot}-\vec{r'}_\bot \mid^2\right]^2} \Bigg]\cr
&& ~~~~~~~~~~~~~~~~~~~~~~~~~~~~~~~\times\int_{0}^{\infty} \frac{
d\xi}{\xi} \exp\left[+i \omega \frac{\mid
\vec{r}_{\bot}-\vec{r'}_\bot \mid^2}{2c \xi} +i k_w \xi\right] \cr
&& ~~~~~+ \left[\frac{ic}{\omega}
\frac{\vec{r}_{\bot}-\vec{r'}_{\bot}} {\mid\vec{r}_{\bot}
-\vec{r'}_{\bot}\mid}\cdot\frac{d}{d\left[\mid\vec{r}_{\bot}-\vec{r'}_{\bot}\mid\right]}\right]\cr
&& ~~~~~~~~~~~~~~~~~~~~~~~~~~~~~~~\times\int_{0}^{\infty} \frac{
d\xi}{\xi} \exp\left[+i \omega \frac{\mid
\vec{r}_{\bot}-\vec{r'}_\bot \mid^2}{2c \xi}- \frac{i\omega \xi}{2
c \bar{\gamma}_z^2}\right] \Bigg\}\label{effp2}
\end{eqnarray}
for the transverse field and

\begin{eqnarray}
{\widetilde{E}}_{z}(z, \vec{r}_{\bot}) &=& - \frac{i \omega
\bar{f}(\omega) }{c} \int d \vec{r'}_{\bot}
{\rho_o}\left(\vec{r'}_\bot\right) \exp\left[\frac{i\omega z}{2 c
\bar{\gamma}_z^2}\right]\cr &&\times \Bigg\{-\exp[i k_w
z]\left[\frac{c K}{2 \omega \gamma}\frac{x-x'}{\mid
\vec{r}_{\bot}-\vec{r'}_\bot \mid}\cdot\frac{d}{d\left[\mid
\vec{r}_{\bot}-\vec{r'}_\bot \mid\right]}\right] \cr &&
~~~~~~\times\int_{0}^{\infty} \frac{d\xi}{\xi} \exp\left[i
\omega\frac{\mid \vec{r}_{\bot}-\vec{r'}_\bot \mid^2}{2c \xi}-ik_w
\xi\right] \cr && ~~~~~~- \exp[-i k_w z]\left[\frac{c K}{2 \omega
\gamma}\frac{x-x'}{\mid \vec{r}_{\bot}-\vec{r'}_\bot
\mid}\cdot\frac{d}{d\left[\mid \vec{r}_{\bot}-\vec{r'}_\bot
\mid\right]}\right] \cr &&  ~~~~~~ \times\int_{0}^{\infty}
\frac{d\xi}{\xi} \exp\left[+i \omega\frac{\mid
\vec{r}_{\bot}-\vec{r'}_\bot \mid^2}{2c \xi}+ i k_w \xi\right]\cr
&& ~~~~~~+\frac{1}{\bar{\gamma}_z^2}\int_{0}^{\infty}
\frac{d\xi}{\xi} \exp\left[+i \omega\frac{\mid
\vec{r}_{\bot}-\vec{r'}_\bot \mid^2}{2c \xi}- \frac{i\omega \xi}{2
c \bar{\gamma}_z^2} \right]\Bigg\}~ \label{effz2}
\end{eqnarray}
for the longitudinal field.

We now use the fact that, for any real number $\alpha>0$:

\begin{eqnarray}
\left\{
\begin{array}{c}  \int_{0}^{\infty} {d \xi}
\exp\left[i\left(-\xi+{\alpha}/{\xi}\right)\right]/{\xi} = 2
K_0\left(2~
\sqrt{\alpha}\right) ~~~~~~~~~~~~~~~~~~~~~~~~~~~~~~~~~~~~~~~~~~~~~~~~~~~ \\
 \int_{0}^{\infty} {d \xi}
\exp\left[i\left(+\xi+{\alpha}/{\xi}\right)\right]/{\xi} = 2
K_0\left(-2 i \sqrt{\alpha}\right)= \pi\left[i J_0(2
\sqrt{\alpha})- Y_0(2\sqrt{\alpha})\right]~,
\end{array} \right. \label{rele}
\end{eqnarray}
where $K_n$ is the n-th order modified Bessel function of the
second kind, $Y_n$ is n-th order  Bessel function of the second
kind and $J_n$ is the n-th order Bessel function of the first
kind. Using Eq. (\ref{rele}) and the fact that $k_w \gg \omega/(2
c \bar{\gamma}_z^2)$, and remembering that

\begin{eqnarray}
&&\frac{d^2}{d \left[\mid \vec{r}_{\bot}-\vec{r'}_\bot
\mid^2\right]^2} = \cr &&\frac{1}{4 \mid
\vec{r}_{\bot}-\vec{r'}_\bot \mid^2} \frac{d^2}{d \left[\mid
\vec{r}_{\bot}-\vec{r'}_\bot \mid\right]^2}-\frac{1}{ 4 \mid
\vec{r}_{\bot}-\vec{r'}_\bot \mid^3}\frac{d}{d\left[\mid
\vec{r}_{\bot}-\vec{r'}_\bot \mid\right]} \label{der}
\end{eqnarray}

we can write Eq. (\ref{effp2}) and Eq. (\ref{effz2}) as

\begin{eqnarray}
&& \vec{\widetilde{E}}_{\bot}(z, \vec{r}_{\bot}) = -\frac{i \omega
\bar{f}(\omega)}{c} \int d \vec{r'}_{\bot}
{\rho_o}\left(\vec{r'}_\bot\right) \exp\left[ \frac{i\omega z}{2 c
\bar{\gamma}_z^2}\right] \cr && \times \Bigg\{+\exp[+i k_w z]
\frac{K \vec{e}_x}{i\gamma} K_0\left(\sqrt{2} \frac{\mid
\vec{r}_{\bot}-\vec{r'}_\bot \mid}{\sqrt{\lambdabar \lambdabar_w}}
\right)  \cr &&~~~~~- \exp[-i k_w z] \frac{K \vec{e}_x}{i\gamma}
K_0\left(-\sqrt{2} i \frac{\mid \vec{r}_{\bot}-\vec{r'}_\bot
\mid}{\sqrt{\lambdabar \lambdabar_w}} \right) \cr&& ~~~~~+\exp[+i
k_w z] \frac{i c r_w}{ \omega} \Bigg[-\frac{\sqrt{2}
\vec{e}_x}{\sqrt{\lambdabar\lambdabar_w}\mid
\vec{r}_{\bot}-\vec{r'}_\bot \mid}K_1\left(\sqrt{2}  \frac{\mid
\vec{r}_{\bot}-\vec{r'}_\bot \mid}{\sqrt{\lambdabar \lambdabar_w}}
\right)\cr &&
~~~~~~~~~~~~~~~~~~~~~~~~~~~~~~~~~~~~+\frac{2(x-x')(\vec{r}_{\bot}-\vec{r'}_\bot)}{\lambdabar
\lambdabar_w \mid \vec{r}_{\bot}-\vec{r'}_\bot \mid^2}
K_2\left(\sqrt{2} \frac{\mid \vec{r}_{\bot}-\vec{r'}_\bot
\mid}{\sqrt{\lambdabar \lambdabar_w}} \right)\Bigg]\cr &&~~~~~-
\exp[-i k_w z] \frac{i c r_w}{ \omega} \Bigg[-\frac{\sqrt{2} i
\vec{e}_x}{\sqrt{\lambdabar\lambdabar_w}\mid
\vec{r}_{\bot}-\vec{r'}_\bot \mid}K_1\left(-\sqrt{2} i \frac{\mid
\vec{r}_{\bot}-\vec{r'}_\bot \mid}{\sqrt{\lambdabar \lambdabar_w}}
\right)\cr &&~~~~~~~~~~~~~~~~~~~~~~~~~~~~~~~~~~~~~~
+\frac{{2}(x-x')(\vec{r}_{\bot}-\vec{r'}_\bot)}{\lambdabar
\lambdabar_w \mid \vec{r}_{\bot}-\vec{r'}_\bot \mid^2}
K_2\left(-\sqrt{2} i \frac{\mid \vec{r}_{\bot}-\vec{r'}_\bot
\mid}{\sqrt{\lambdabar \lambdabar_w}} \right)\Bigg]  \cr &&~~~~~
-\left[\frac{ic}{\omega}\frac{\vec{r}_{\bot}-\vec{r'}_{\bot}}{\mid\vec{r}_{\bot}-\vec{r'}_{\bot}\mid}\right]
\frac{2}{\bar{\gamma}_z \lambdabar} K_1\left( \frac{\mid
\vec{r}_{\bot}-\vec{r'}_\bot \mid}{\bar{\gamma}_z \lambdabar}
\right) \Bigg\} \label{Teffp3}
\end{eqnarray}
and

\begin{eqnarray}
{\widetilde{E}}_{z}(z, \vec{r}_{\bot}) &=& - \frac{i \omega
\bar{f}(\omega) }{c} \int d \vec{r'}_{\bot}
{\rho_o}\left(\vec{r'}_\bot\right) \exp\left[\frac{i\omega z}{2 c
\bar{\gamma}_z^2}\right]\cr &&\times
\Bigg\{+\frac{\sqrt{2}}{\sqrt{\lambdabar \lambdabar_w}} \exp[+i
k_w z]\left[\frac{c K}{\omega \gamma}\frac{x-x'}{\mid
\vec{r}_{\bot}-\vec{r'}_\bot \mid}\right] K_1\left(\sqrt{2}~
\frac{\mid \vec{r}_{\bot}-\vec{r'}_\bot \mid}{\sqrt{\lambdabar
\lambdabar_w}} \right)\cr &&
~~~~~+\frac{\sqrt{2}i}{\sqrt{\lambdabar \lambdabar_w}}  \exp[-i
k_w z]\left[\frac{c K}{\omega \gamma}\frac{x-x'}{\mid
\vec{r}_{\bot}-\vec{r'}_\bot \mid}\right] K_1\left(\sqrt{2}
i\frac{\mid \vec{r}_{\bot}-\vec{r'}_\bot \mid}{\sqrt{\lambdabar
\lambdabar_w}} \right)\cr && ~~~~~+\frac{2}{\bar{\gamma}_z^2}
K_0\left( \frac{\mid \vec{r}_{\bot}-\vec{r'}_\bot
\mid}{\bar{\gamma}_z \lambdabar} \right) \Bigg\}~. \label{Teffz3}
\end{eqnarray}
Note that, similarly as in Eq. (\ref{effp}) Eq. (\ref{effz}), it
is possible to recognize in Eq. (\ref{Teffp3}) and Eq.
(\ref{Teffz3}) radiative and space-charge terms, as well as
gradient and current terms (for the transverse field components).

\subsection{\label{sub:gauss} Cross-check with Gauss law}

It is possible to cross-check our expressions for the field with
the help of Gauss law:

\begin{equation}
\vec{\nabla} \cdot \vec{\bar{E}} = 4\pi \bar{\rho} ~.
\label{gauss}
\end{equation}
This cross-check will constitute a general cross-check of the
correctness of our calculation and allow a better understanding of
the interplay of different field contributions in the complicated
machinery of Maxwell's equation.

As said before, in the present study we work in the steady state,
when the bunch has travelled inside the undulator for more than
$2\bar{\gamma}_z^2 \lambdabar$. In this case, an explicit
expression for transverse and longitudinal fields are given in Eq.
(\ref{Teffp3}) and Eq. (\ref{Teffz3}).

We will demonstrate that the field $\vec{\bar{E}} =
\vec{\bar{E}}_{\mathrm{rad}} + \vec{\bar{E}}_{\mathrm{sc}}$ obeys
Eq. (\ref{gauss}) by separately showing that radiation field
$\vec{\bar{E}}_{\mathrm{rad}}$ and space-charge field
$\vec{\bar{E}}_{\mathrm{sc}}$ verify:

\begin{eqnarray}
&&\vec{\nabla} \cdot \vec{\bar{E}}_{\mathrm{sc}} = 4\pi
\bar{\rho}\cr && \vec{\nabla} \cdot \vec{\bar{E}}_{\mathrm{rad}} =
0 ~. \label{gauss2}
\end{eqnarray}
Relations (\ref{gauss2}) can be interpreted saying that the
radiation field is not entangled with sources, while the
space-charge field is. Hence the different right hand sides.

Let us begin with the space-charge field. First we can write:

\begin{equation}
\frac{\partial \bar{E}_{z~\mathrm{sc}}}{\partial z} =
\frac{\partial}{\partial z} \left\{\widetilde{E}_{z~\mathrm{sc}}
\exp[i\omega z/c] \right\} =
\frac{i\omega}{c}\left(1+\frac{1}{2\bar{\gamma}_z}^2\right)
\bar{E}_{z\mathrm{sc}} \simeq \frac{i\omega}{c}
\bar{E}_{z\mathrm{sc}}~, \label{dez}
\end{equation}
because, in the steady state, $\widetilde{E}_{z~\mathrm{sc}}$
depends on $z$ only through $\exp[i \omega z/(2 c
\bar{\gamma}_z^2]$.

Thus, in order to verify the first of Eq. (\ref{gauss2}) we should
prove that

\begin{eqnarray}
&&\vec{\nabla}_\bot \cdot \vec{\bar{E}}_{\bot\mathrm{sc}} = 4\pi
\bar{\rho} - \frac{i\omega}{c} \bar{E}_{z\mathrm{sc}}~.
\label{gauss2sc}
\end{eqnarray}
From Eq. (\ref{Teffp3}) we have

\begin{eqnarray}
&& \vec{\widetilde{E}}_{\bot\mathrm{sc}} = -\frac{2
\bar{f}(\omega) }{\bar{\gamma}_z \lambdabar}\exp\left[
\frac{i\omega z}{2 c \bar{\gamma}_z^2}\right] \int d
\vec{r'}_{\bot} {\rho_o}\left(\vec{r'}_\bot\right)
\frac{\vec{r}_{\bot}-\vec{r'}_{\bot}}{\mid\vec{r}_{\bot}-\vec{r'}_{\bot}\mid}
 K_1\left( \frac{\mid
\vec{r}_{\bot}-\vec{r'}_\bot \mid}{\bar{\gamma}_z \lambdabar}
\right)~.\cr && \label{Teffp3sc}
\end{eqnarray}
In order to calculate the left hand side of Eq. (\ref{gauss2sc})
we can use the divergence theorem in two dimensions, to find:

\begin{eqnarray}
\vec{\nabla}_\bot\cdot\left\{
\frac{\vec{R}_\bot}{\left|\vec{R}_\bot\right|}
K_1\left(\frac{\omega
\left|\vec{R}_\bot\right|}{c\bar{\gamma}_z}\right)\right\}=-\frac{\omega}{c\bar{\gamma}_z}
K_o\left(\frac{\omega
\left|\vec{R}_\bot\right|}{c\bar{\gamma_z}}\right) +\frac{2\pi c
\bar{\gamma_z}}{\omega} \delta\left(\vec{R}_\bot\right) ~,
\label{nabezbar}
\end{eqnarray}
where $\delta$ indicates the Dirac-delta function and derivation
is understood in weak sense, we set $\vec{R}_\bot=\vec{r}_\bot
-\vec{r'}_\bot$. Remembering $\vec{\widetilde{E}}_\bot
=\vec{\bar{E}}_\bot \exp[-i \omega z/c]$ we obtain

\begin{eqnarray}
\vec{\nabla}_\bot \cdot \vec{\bar{E}}_{\bot \mathrm{sc}}&=& \left[
- \frac{2\omega^2}{c^2\bar{\gamma}_z^2} \int d\vec{r'}_\bot
\rho_o(\vec{r'}_\bot) K_o\left(\frac{
\left|\vec{r}_\bot-\vec{r'}_\bot\right|}{\lambdabar
\bar{\gamma}_z}\right) + 4\pi \rho_o(\vec{r}_\bot) \right] \cr
&&\times \bar{f}(\omega) \exp\left[\frac{i\omega z }{v_z}\right]
~. \label{ezbar3}
\end{eqnarray}
Now, from Eq. (\ref{Teffz3}) we have:

\begin{eqnarray}
{\bar{E}}_{z~\mathrm{sc}}&=& - \frac{2i \omega  }{c
\bar{\gamma}_z^2}  \int d \vec{r'}_{\bot}
{\rho_o}\left(\vec{r'}_\bot\right) \exp\left[\frac{i\omega z}{2 c
\bar{\gamma}_z^2}\right]  K_0\left( \frac{\mid
\vec{r}_{\bot}-\vec{r'}_\bot \mid}{\lambdabar \bar{\gamma}_z
}\right) \bar{f}(\omega) \exp\left[\frac{i\omega z }{v_z}\right]
~.\cr && \label{Teffz333}
\end{eqnarray}
Substitution in the right hand side of Eq. (\ref{gauss2sc}) yields
Eq. (\ref{ezbar3}), thus verifying Eq. (\ref{gauss2sc}).

Let us now consider the radiative fields, and show that the second
of Eq. (\ref{gauss2}) is also verified. Presentations in Eq.
(\ref{Teffp3}) and Eq. (\ref{Teffz3}) include now many terms, and
it is convenient to start with alternative presentations for the
transverse and longitudinal field, namely Eq. (\ref{perpexp4}) and
Eq. (\ref{longexp4}). In the limit for $z\gg \bar{\gamma}_z^2
\lambdabar$ The radiative part of the field is given by:

\begin{eqnarray}
&&\vec{\bar{E}}_{\bot\mathrm{rad}}(z, \vec{r}_{\bot}) =
\exp\left[\frac{i\omega z}{c}\right] \frac{\omega^2
\bar{f}(\omega)}{2\pi c^2} \int d\vec{\theta} \int_{0}^{\infty}
dz' \int d \vec{r'}_{\bot} {\rho_o}\left(\vec{r'}_\bot\right) \cr
&&\times \exp\left[-\frac{i \omega}{c} \vec{\theta} \cdot \left(
\vec{r}_\bot -\vec{r'}_\bot\right)\right] \exp\left[\frac{i \omega
\theta^2 (z'-z)}{2 c }\right]\cr && \times \left\{ \frac{K
\vec{e}_x}{\gamma} \sin[k_w z']-\frac{i \omega \theta_x
r_w\vec{\theta}}{c} \cos[k_w z']\right\} \label{pexp}
\end{eqnarray}
and

\begin{eqnarray}
&&{\bar{E}}_{z~\mathrm{rad}}(z, \vec{r}_{\bot}) = - \frac{\omega^2
\bar{f}(\omega) K }{2\pi c^2 \gamma}\exp\left[\frac{i\omega
z}{c}\right]    \int d\vec{\theta} \int_{0}^{\infty} dz' \int d
\vec{r'}_{\bot}
{\rho_o}\left(\vec{r'}_\bot\right)\cr&&\times\exp\left[-\frac{i
\omega}{c} \vec{\theta} \cdot \left( \vec{r}_\bot -\vec{r'}_\bot
\right)\right] \exp\left[\frac{i \omega \theta^2 (z'-z)}{2 c
}\right] \theta_x \sin[k_w z']~. \label{gexp}
\end{eqnarray}
Here we neglected factors $\exp[i \omega z'/(2 c
\bar{\gamma}_z^2)]$  because $\omega \lambda_w / (2 c
\bar{\gamma}_z^2) \ll 1$. We may now calculate directly
$\vec{\nabla}_\bot \cdot\bar{E}_{\bot\mathrm{rad}}$ and
$\partial_z {\bar{E}}_{z~\mathrm{rad}}$. We obtain

\begin{eqnarray}
&&\vec{\nabla}_\bot \cdot \vec{\bar{E}}_{\bot\mathrm{rad}}(z,
\vec{r}_{\bot}) =-\frac{i\omega}{c}  \frac{K \omega^2
\bar{f}(\omega)}{2\pi c^2 \gamma} \exp\left[\frac{i\omega
z}{c}\right] \int d\vec{\theta} \int_{0}^{\infty} dz' \int d
\vec{r'}_{\bot} {\rho_o}\left(\vec{r'}_\bot\right) \cr &&\times
\exp\left[-\frac{i \omega}{c} \vec{\theta} \cdot \left(
\vec{r}_\bot -\vec{r'}_\bot\right)\right] \Bigg\{ { \theta_x}
\sin[k_w z']\exp\left[\frac{i \omega \theta^2 (z'-z)}{2 c }\right]
\cr &&  -{2 \theta_x \lambdabar_w }\cos[k_w z']
\frac{\partial}{\partial z'} \exp\left[\frac{i \omega \theta^2
(z'-z)}{2 c }\right] \Bigg\} \label{divper}
\end{eqnarray}
and

\begin{eqnarray}
&&\partial_z {\bar{E}}_{z~\mathrm{rad}}(z, \vec{r}_{\bot}) =
-\frac{i \omega}{c} \frac{K \omega^2 \bar{f}(\omega) }{2\pi c^2
\gamma}\exp\left[\frac{i\omega z}{c}\right]    \int d\vec{\theta}
\int_{0}^{\infty} dz' \int d \vec{r'}_{\bot}
{\rho_o}\left(\vec{r'}_\bot\right)\cr&&\times\exp\left[-\frac{i
\omega}{c} \vec{\theta} \cdot \left( \vec{r}_\bot -\vec{r'}_\bot
\right)\right] \exp\left[\frac{i \omega \theta^2 (z'-z)}{2 c
}\right]\cdot  \left\{\theta_x \sin[k_w z'] \right\}~,
\label{divlon}
\end{eqnarray}
In Eq. (\ref{divlon}) we neglected an extra-term $-
{\theta^2}\theta_x/2 \sin[k_w z']$ in parenthesis $\{...\}$,
because $\lambdabar/\lambdabar_w \ll 1$. We now integrate by parts
the term in $\cos[k_w z']$  in Eq. (\ref{divper}) to obtain

\begin{eqnarray}
&&\vec{\nabla}_\bot \cdot \vec{\bar{E}}_{\bot\mathrm{rad}}(z,
\vec{r}_{\bot}) =-\frac{i\omega}{c}  \frac{K \omega^2
\bar{f}(\omega)}{2\pi c^2 \gamma} \exp\left[\frac{i\omega
z}{c}\right] \int d\vec{\theta} \int_{0}^{\infty} dz' \int d
\vec{r'}_{\bot} {\rho_o}\left(\vec{r'}_\bot\right) \cr &&\times
\exp\left[-\frac{i \omega}{c} \vec{\theta} \cdot \left(
\vec{r}_\bot -\vec{r'}_\bot\right)\right] \exp\left[\frac{i \omega
\theta^2 (z'-z)}{2 c }\right]\exp\left[ \frac{i\omega z'}{2 c
\bar{\gamma}_z^2} \right] \cdot \left\{ -   \theta_x \sin[k_w
z']\right\}~.\cr &&\label{divper2}
\end{eqnarray}
Obviously $\vec{\nabla}_\bot \cdot
\vec{\bar{E}}_{\bot\mathrm{rad}}+\partial_z
{\bar{E}}_{z~\mathrm{rad}}=0$, and also the second of Eq.
(\ref{gauss2}) is satisfied.

\section{\label{sub:wake} Wake function and impedance}

The knowledge of the electric field at the position of any test
particle inside the beam, that has been derived in Section
(\ref{sec:main}), allows to derive wake function and impedance
related with the system under study. Let us briefly review here
concepts of longitudinal impedance and wake function, that will be
used later on. The longitudinal impedance of a system,
$Z_o(\omega)$, can be given as the Fourier transform of the wake
function $G_o(\Delta s)$:

\begin{eqnarray}
&&Z_o = \int_{-\infty}^{\infty} \frac{d(\Delta s)}{\beta c}~
G_o(\Delta s) \exp\left[i \omega \frac{\Delta s}{\beta
c}\right]~,\cr && G_o = \frac{1}{(-e)} \int_{-\infty}^{\infty} d
\vec{r}' \cdot ~\vec{E}^{o}(\Delta s, \vec{r'}(t), t)|_{ t =
z'/(\beta_z c)}~. \label{impe}
\end{eqnarray}
Here the integral in the expression for $G_o$ is a line integral
calculated along the trajectory of a test particle. In fact,
$\vec{E}^{o}(\Delta s, \vec{r'}(t), t)$ indicates the time-domain
electric field generated by a source particle acting on the test
particle at longitudinal distance $\Delta s$ from the source. In
the calculation of $\vec{E}^{o}(\Delta s, \vec{r'}(t), t)$ we
assume that effects from the vacuum chamber are negligible, i.e.
$a \gg \bar{\gamma}_z \lambdabar$, that is the third of conditions
(\ref{cons1}). $\vec{E}^{o}(\Delta s, \vec{r'}(t), t)$ is
integrated along the test particle trajectory, and divided by the
electron charge $(-e)$, so that $e^2 G_o(\Delta s)$ is the energy
(gained, or lost) by the test particle due to the action of the
source. In agreement with \cite{WAKK} we take the test particle
behind the source for positive values of $\Delta s$.

According to the given definition of wake function, one should
integrate the field over the entire trajectory. However, there is
no principle difficulty in considering only part of the
trajectory, let us say, up to longitudinal position $z$.
Mathematically, this means that the line integral for $G_o$ should
be performed up to the trajectory point of the test electron
corresponding to longitudinal position $z$. In this way,
$G=G(\Delta s, z)$.

Note that Eq. (\ref{impe}) is automatically dependent on a
particular source electron, and a particular test electron. In
order to formulate this statement in a mathematical way, we may
introduce test and source particle initial transverse position
$\vec{r}_{\bot T}$ and $\vec{r}_{\bot S}$ and write $G_o =
G_o(\Delta s,z,\vec{r}_{\bot S},\vec{r}_{\bot T})$, where we
neglect differences in energy between the two particles. Following
our previous work \cite{OURL}, we will slightly modify the
concepts of wake and impedance by substituting test and source
particles with disks of total charge (-e), longitudinally
separated by a distance $\Delta s$. This amounts to an integration
over the transverse particle distribution in $d\vec{r}_{\bot T}$
and $d\vec{r}_{\bot S}$, that makes our definitions independent of
$\vec{r}_{\bot T}$ and $\vec{r}_{\bot S}$.  We thus obtain

\begin{eqnarray}
G(\Delta s,z) = c^2 \int d \vec{r'}_\bot \int d \vec{r''}_\bot
\rho_o(\vec{r'}_\bot) \rho_o(\vec{r''}_\bot) G_o(\Delta
s,z,\vec{r'}_{\bot},\vec{r''}_{\bot})~. \label{Gred}
\end{eqnarray}
In Eq. (\ref{Gred}) we used the fact that $\rho_o$ is independent
of the longitudinal position. With the redefinition in Eq.
(\ref{Gred}) we can further consider the impedance $Z(\omega,z)$
proceeding as in (\ref{impe}) for the definition of $Z_o(\omega)$,
but Fourier substituting $G_o(\Delta s)$ with $G(\Delta s,z)$.
Note that by definition of $Z(\omega,z)$ we have

\begin{eqnarray}
Z(\omega,z) = \frac{1}{|\bar{f}(\omega)|^2} \int_V
\vec{{\bar{j}}}^{~*} \cdot {\vec{\bar{E}}} ~d V =
\frac{1}{|\bar{f}(\omega)|^2} \int_{0}^z dz' \int_A d\vec{r'}_\bot
\vec{{\bar{j}}}^{~*} \cdot {\vec{\bar{E}}} ~ ,\label{defZZ}
\end{eqnarray}
where $|\bar{f}(\omega)|^{-2}$ accounts for the fact that test and
source disks have total charge (-e), while $\vec{{\bar{j}}}^{~*}
\cdot {\vec{\bar{E}}}\propto |\bar{f}(\omega)|^2$,
$\bar{f}(\omega)$ being the already defined Fourier transform of
the longitudinal bunch-profile. Here the volume $V$ is a cylinder
of base A including the undulator up to position $z'=z$. The
integration in $z'$ is performed from $0$ to $z$, because we will
be interested in impedance and wakes generated inside the
undulator, and we will assume that the undulator begins at
position $z=0$.

\section{\label{sec:impe} Impedance calculations}

According to Eq. (\ref{defZZ}), the expressions for the
longitudinal impedance associated with
$\vec{\widetilde{E}}_{\bot}$ and $\vec{\widetilde{E}}_{z}$,
${Z}_\bot$ and ${Z}_z$ are given by

\begin{eqnarray}
{Z}_\bot &=& \frac{1}{|\bar{f}(\omega)|^2} \int_0^z dz'
\exp\left[-\frac{i s_o(z')}{v_o}\right] \cr &&\times \int
d\vec{r'}_\bot \bar{f}^*(\omega) \rho_o^*[\vec{r'}_\bot- r_w
\cos(k_w z)] \vec{v}_{o\bot}(z')\cdot
\vec{\bar{E}}_\bot(z',\omega,\vec{r'}_\bot) \cr  Z_z &=&
\frac{1}{|\bar{f}(\omega)|^2} \int_0^z dz' \exp\left[-\frac{i
s_o(z')}{v_o}\right] \cr &&\times   \int d\vec{r'}_\bot
\bar{f}^*(\omega) \rho_o^*[\vec{r'}_\bot- r_w \cos(k_w
z)\vec{e}_x] c {\bar{E}}_z(z',\omega,\vec{r'}_\bot) ~.\cr &&
\label{imped0}
\end{eqnarray}
When we calculated the field, we saw that the dependence of the
source transverse charge density $\rho_o$ on the electron motion
$\vec{r'}_{\bot o}(z)$ had to be accounted for. We will show that,
in the calculation of the impedance, the dependence of the test
transverse charge density $\rho_o$ on $\vec{r'}_\bot- r_w \cos(k_w
z)\vec{e}_x$  must also be accounted for. With the help of a
change of variables we rewrite Eq. (\ref{imped0}) as a sum
$Z=Z_\bot+Z_z$:

\begin{eqnarray}
{Z} &=& \frac{1}{|\bar{f}(\omega)|^2} \int_0^z dz'
\exp\left[-\frac{i s_o(z')}{v_o}\right] \int d\vec{r'}_\bot
\bar{f}^*(\omega) \rho_o^*(\vec{r'}_\bot) \cr && \times
\Bigg\{-\frac{K c}{2 i \gamma} \left[\exp(i k_w z')-\exp(-i k_w
z') \right] {\bar{E}}_x(z',\omega,\vec{r'}_\bot+r_w \cos(k_w
z)\vec{e}_x) \cr &&~~~~~+ c
{\bar{E}}_z(z',\omega,\vec{r'}_\bot+r_w \cos(k_w
z)\vec{e}_x)\Bigg\} ~.\cr && \label{imped}
\end{eqnarray}
Calculations can be drastically simplified, because transverse
radiative gradient terms in $Z_\bot$ cancel with longitudinal
radiative terms in $Z_z$. It is easier to show this facts with the
help of Eq. (\ref{effp}) and Eq. (\ref{effz}), rather than using
explicit expressions Eq. (\ref{Teffp3}) and Eq. (\ref{Teffz3}).

First, with the help of Eq. (\ref{effp}) and Eq. (\ref{effz}), we
write down the part of the impedance from the radiative transverse
field, $Z_{\bot \mathrm{r}}$ and from the radiative longitudinal
field, $Z_{z\mathrm{~r}}$. In principle, in order to dispose of
the oscillating terms in
$\vec{\bar{E}}_\mathrm{r}(z',\omega,\vec{r'}_\bot+r_w \cos(k_w
z)\vec{e}_x)$, we may use the same mathematical shortcut that can
be exploited to obtain Eq. (\ref{effp}) and Eq. (\ref{effz}) from
Eq. (\ref{generalfin2}) and Eq. (\ref{generalfinz2b}). In fact, we
may formally expand the Green's function exponential $\exp\{i
\omega [{\mid \vec{r'}_{\bot}-\vec{r''}_\bot +r_w \cos(k_w z')
\vec{e}_x\mid^2}/{2c (z'-z'')}]\}$ to the first order in $r_w$ and
keep non-negligible first-harmonic terms in $\exp[\pm i k_w z']$.
However, $\vec{\bar{E}}_{\bot \mathrm{r}}$ is multiplied by
$\vec{v}_{o\bot}$ in the expression for the impedance $Z_\bot$.
Since $\vec{v}_{o\bot}$ oscillates with period $\lambdabar_w$, we
can neglect the oscillatory contributions in $\cos(k_w z)
\vec{e}_x$ in the expression of $\vec{\bar{E}}_{\bot\mathrm{r}}$,
because they would give oscillatory contributions that average to
zero after integration in $dz'$. We therefore obtain:

\begin{eqnarray}
&& {Z}_{\bot\mathrm{r}} = \frac{K \omega}{2  \gamma} \int
d\vec{r'}_\bot \int d \vec{r''}_{\bot} \rho_o^*(\vec{r'}_\bot)
{\rho_o}\left(\vec{r''}_\bot\right) \int_0^z dz'  \cr && \times
\Bigg\{- \int_{0}^{z'} \frac{ dz''}{z'-z''} \exp\left[i \omega
\frac{\mid \vec{r'}_{\bot}-\vec{r''}_\bot \mid^2}{2c (z'-z'')}
\right] \left[+\frac{K \vec{e}_x}{2i\gamma}\exp[ik_w
(z''-z')]\right] \cr &&~~~~~+\int_{0}^{z'} \frac{ dz''}{z'-z''}
\exp\left[i \omega \frac{\mid \vec{r'}_{\bot}-\vec{r''}_\bot
\mid^2}{2c (z'-z'')} \right] \left[-\frac{K
\vec{e}_x}{2i\gamma}\exp[ik_w(z'- z'')] \right] \cr && ~~~~~+
\int_{0}^{z'} \frac{ dz''}{z'-z''} \exp\left[i \omega \frac{\mid
\vec{r'}_{\bot}-\vec{r''}_\bot \mid^2}{2c (z'-z'')} \right] \cr &&
~~~~~~~~~~~~~~~~~~~\times\left[ +\frac{r_w}{2 (z'-z'')} +
\frac{i\omega r_w(x'-x'')^2}{2 c(z'-z'')^2} \right]\exp[ik_w
(z''-z')] \cr &&~~~~~+ \int_{0}^{z'} \frac{ dz''}{z'-z''}
\exp\left[i \omega \frac{\mid \vec{r'}_{\bot}-\vec{r''}_\bot
\mid^2}{2c (z'-z'')} \right]\cr && ~~~~~~~~~~~~~~~~~~~\times\left[
-\frac{r_w }{2 (z'-z'')} - \frac{i\omega r_w(x'-x'')^2}{2
c(z'-z'')^2} \right]\exp[ik_w (z'-z'')]  ~.\cr && \label{effpr}
\end{eqnarray}
for the transverse field. On the contrary, the longitudinal
velocity $\vec{v}_{o~z}$ is a sum of a constant term, whose
magnitude is about $c$ and a negligible oscillates with period
$2\lambdabar_w$. As a result, the oscillatory contributions in
$\cos(k_w z) \vec{e}_x$ in the expression of
${\bar{E}}_{z~\mathrm{r}}$ must be kept, and an expansion of the
exponential in the Green's function must be performed, leading to

\begin{eqnarray}
&&Z_{z~\mathrm{r}} =  -\frac{K \omega }{2 \gamma}\int
d\vec{r'}_\bot \int d \vec{r''}_{\bot} \rho_o^*(\vec{r'}_\bot)
{\rho_o}\left(\vec{r''}_\bot\right) \int_0^z dz' \cr && \times
\Bigg\{\int_{0}^{z'} \frac{dz''}{z'-z''} \exp\left[i
\omega\frac{\mid \vec{r'}_{\bot}-\vec{r''}_\bot \mid^2}{2c
(z'-z'')}\right]\cr &&~~~~~~~~~~~~~~~~~~~\times \left[ + \frac{i
\omega r_w (x'-x'')^2}{2c(z'-z'')^2}+\frac{r_w}{2(z'-z'')}\right]
\exp[i k_w (z''-z')]\cr && + \int_{0}^{z'} \frac{dz''}{z'-z''}
\exp\left[i \omega\frac{\mid \vec{r'}_{\bot}-\vec{r''}_\bot
\mid^2}{2c (z'-z'')}\right]\cr &&~~~~~~~~~~~~~~~~~~~ \times\left[
-\frac{i\omega r_w (x'-x'')^2}{2 c(z'-z'')^2
}-\frac{r_w}{2(z'-z'') }\right]\exp[i k_w (z'-z'')]\Bigg\}~
\label{effzr}
\end{eqnarray}
Partial cancellation can be exploited between the last two terms
of Eq. (\ref{effpr}) and Eq. (\ref{effzr}). Note that the
longitudinal radiative impedance is completely cancelled, and one
obtains

\begin{eqnarray}
&&Z_\mathrm{r}={Z}_{\bot{r}} + {Z}_{z\mathrm{~r}}= \cr && \frac{K
\omega}{2  \gamma} \int d\vec{r'}_\bot \int d \vec{r''}_{\bot}
\rho_o^*(\vec{r'}_\bot) {\rho_o}\left(\vec{r''}_\bot\right)
\int_0^z dz'  \cr && \times \Bigg\{- \int_{0}^{z'} \frac{
dz''}{z'-z''} \exp\left[i \omega \frac{\mid
\vec{r'}_{\bot}-\vec{r''}_\bot \mid^2}{2c (z'-z'')} \right]
\left[+\frac{K }{2i\gamma}\exp[ik_w (z''-z')]\right] \cr
&&~~~~~+\int_{0}^{z'} \frac{ dz''}{z'-z''} \exp\left[i \omega
\frac{\mid \vec{r'}_{\bot}-\vec{r''}_\bot \mid^2}{2c (z'-z'')}
\right] \left[-\frac{K }{2i\gamma}\exp[ik_w(z'- z'')] \right] ~.
\label{Zrad}
\end{eqnarray}
Performing the integrals in $dz''$ with the help of Eq.
(\ref{rele}) or, equivalently, calculating $Z_\mathrm{r}$  with
the help of the transverse radiative current terms in Eq.
(\ref{Teffp3}) we finally obtain

\begin{eqnarray}
&&Z_\mathrm{r}={Z}_{\bot{r}} + {Z}_{z\mathrm{~r}}=  i\frac{K^2
\omega z}{2   \gamma^2} \int d\vec{r'}_\bot \int d
\vec{r''}_{\bot} \rho_o^*(\vec{r'}_\bot)
{\rho_o}\left(\vec{r''}_\bot\right) \cr && \times \Bigg\{
K_o\left[\frac{\sqrt{2}\mid \vec{r'}_{\bot}-\vec{r''}_\bot \mid}
{\sqrt{\lambdabar\lambdabar_w}}\right]+
K_o\left[\frac{-\sqrt{2}i\mid \vec{r'}_{\bot}-\vec{r''}_\bot \mid}
{\sqrt{\lambdabar\lambdabar_w}}\right] \Bigg\}~. \label{Zradfin}
\end{eqnarray}
Let us now consider the space-charge part of the impedance. One
finds that the space-charge term in $Z_\bot$, i.e.
$Z_{\bot\mathrm{sc}}$, averages to zero, as can directly be seen
by inspecting the last term in
$K_1[|\vec{r}_{\bot}-\vec{r'}_\bot|/(\bar{\gamma}_z \lambdabar)]$
of Eq. (\ref{Teffp3}). In fact, such term is independent of $z$.
Now, according to Eq. (\ref{imped}), in order to obtain the
correspondent impedance contribution, this term must be multiplied
by $\vec{v}_\bot$ (i.e. by $\exp[\pm i k_w z']$) and integrated in
$dz'$ for a saturation length. It follows that, during the
integration process in $dz'$, one integrates a fast varying
function of $z$ on the scale of $\lambdabar_w$. As a result, we
obtain a negligible effective impedance contribution over many
undulator periods, and we can neglect the transverse space-charge
contribution in the calculation of wake and impedance. The total
space-charge part of the impedance coincides with the longitudinal
space-charge impedance. It can be shown that oscillatory
contributions in $\cos(k_w z') \vec{e}_x$ in the expression of
$\vec{\bar{E}}_{z\mathrm{sc}}$ are of higher order in
$\omega/\omega_r$ and they can thus be neglected. As a result we
obtain

\begin{eqnarray}
&&Z_\mathrm{sc}={Z}_{z~{sc}} =  -{i \omega
z}\frac{2+{K^2}}{\gamma^2}\int d\vec{r'}_\bot \int d
\vec{r''}_{\bot} \rho_o^*(\vec{r'}_\bot)
{\rho_o}\left(\vec{r''}_\bot\right)  K_0\left( \frac{\mid
\vec{r'}_{\bot}-\vec{r''}_\bot \mid}{\bar{\gamma}_z \lambdabar}
\right) ~.\cr &&\label{Zsc}
\end{eqnarray}
We thus reach the conclusion that only longitudinal space-charge
terms and transverse radiative terms enter the expression for the
impedance that can now be calculated in integral from for any
transverse beam-distribution under conditions (\ref{cons1}).

Finally, we obtain the total impedance
${Z}=Z_{\mathrm{r}}+Z_\mathrm{sc}$. The real part $Z_R$ is given
by

\begin{eqnarray}
&&{Z}_R = -\frac{K^2 \pi \omega z}{4\gamma^2} \int d\vec{r'}_\bot
\int d\vec{r''}_\bot \rho_o^*(\vec{r'}_\bot)
{\rho_o}\left(\vec{r''}_\bot\right)
J_0\left(\frac{\sqrt{2}\left|\vec{r'}_\bot
-\vec{r''}_\bot\right|}{\sqrt{\lambdabar \lambdabar_w}}\right)
~.\label{ZR}
\end{eqnarray}
The imaginary part $Z_I$, instead, amounts to

\begin{eqnarray}
{Z}_I &=& -\frac{K^2 \omega z}{2 \gamma^2} \int d\vec{r'}_\bot
\int d\vec{r''}_\bot \rho_o^*(\vec{r'}_\bot)
{\rho_o}\left(\vec{r''}_\bot\right)  \Bigg\{
\frac{\pi}{2}Y_0\left(\frac{\sqrt{2}\left|\vec{r'}_\bot
-\vec{r''}_\bot\right|}{\sqrt{\lambdabar \lambdabar_w}}\right)\cr
&& - K_0\left(\frac{\sqrt{2}\left|\vec{r'}_\bot
-\vec{r''}_\bot\right|}{\sqrt{\lambdabar \lambdabar_w}}\right)+
\frac{4+2K^2}{K^2} K_0\left(\frac{\left|\vec{r'}_\bot
-\vec{r''}_\bot\right|}{\bar{\gamma}_z \lambdabar }\right)
\Bigg\}~, \label{IR}
\end{eqnarray}
having used the fact that $K_0(-i x) = (\pi/2) [i J_0(x) -
Y_0(x)]$.

\subsection{\label{sub:smalll} Asymptotic case for $\sigma_\bot^2 \ll \lambdabar
\lambdabar_w$}

Before proceeding with the analysis of the wake, it is interesting
to derive asymptotic limits of Eq. (\ref{ZR}) and Eq. (\ref{IR})
in the case for $\sigma_\bot^2 \ll \lambdabar \lambdabar_w$.
Bessel functions in Eq. (\ref{ZR}) and Eq. (\ref{IR}) can be
expanded for small argument values. In particular, using $J_0(x)
\simeq 1$ for $x \ll 1$, and recalling that $\rho_o$ is normalized
to $1/c$, the real part of the impedance becomes

\begin{eqnarray}
{Z}_R &=& - \frac{K^2 \omega \pi z}{4 \gamma^2} \int
d\vec{r'}_\bot \int d\vec{r''}_\bot \rho_o^*(\vec{r'}_\bot)
{\rho_o}\left(\vec{r''}_\bot\right)  = - \frac{K^2  \pi z}{4 c
\lambdabar \gamma^2} ~,\cr && \label{ZRR}
\end{eqnarray}
independently of the choice of $\rho_o$. Subsequently, we use
$K_0(x) \simeq -\gamma_E - \ln(x/2)$ and $Y_0 \simeq 2/\pi
[\gamma_E + \ln(x/2)]$, $\gamma_E\simeq 0.577216$ being the Euler
Gamma constant in the imaginary part of the impedance, Eq.
(\ref{IR}). We obtain

\begin{eqnarray}
{Z}_I &=& -\frac{K^2  \omega z}{\gamma^2} \int d\vec{r'}_\bot \int
d\vec{r''}_\bot \rho_o^*(\vec{r'}_\bot)
{\rho_o}\left(\vec{r''}_\bot\right) \cr && \times \left\{
\ln\left(\sqrt{\frac{\lambdabar}{\lambdabar_r}}\right)-\frac{2}{
K^2} \ln\left(\sqrt{1+\frac{K^2}{2}}\right) -\frac{2
\gamma_E}{K^2}- \frac{2}{K^2}\ln\left(\frac{\left|\vec{r'}_\bot
-\vec{r''}_\bot\right|}{2\lambdabar \gamma}\right)\right\}\cr && =
- \frac{K^2  z}{ c \lambdabar \gamma^2}
\ln\left(\sqrt{\frac{\lambdabar}{\lambdabar_r}}\right)+ \frac{2
z}{ c \lambdabar \gamma^2} \ln\left(\sqrt{1+\frac{K^2}{2}}\right)
+ Z_{I\mathrm{~free}}~, \label{IRR2}
\end{eqnarray}
where

\begin{eqnarray}
Z_{I\mathrm{~free}} &=& \frac{2 z \gamma_E}{c \lambdabar
\gamma^2}+\frac{2 \omega z}{\gamma^2} \int d\vec{r'}_\bot \int
d\vec{r''}_\bot \rho_o^*(\vec{r'}_\bot)
{\rho_o}\left(\vec{r''}_\bot\right)
\ln\left(\frac{\left|\vec{r'}_\bot
-\vec{r''}_\bot\right|}{2\lambdabar \gamma}\right)  ~.\cr
&&\label{ZIFREE}
\end{eqnarray}
$Z_{I\mathrm{~free}}$ is the only model-part of the impedance. In
particular, assuming a Gaussian transverse profile:

\begin{eqnarray}
\rho_o(r_\bot) = \frac{1}{2 \pi \sigma_\bot^2 c}
\exp\left[-\frac{r_\bot^2}{2\sigma_\bot^2}\right] ~\label{rhoog}
\end{eqnarray}
we obtain

\begin{eqnarray}
Z_{I\mathrm{~free}} &=& \frac{2 z \gamma_E}{c \lambdabar
\gamma^2}+\frac{  z}{2 \pi^2 \sigma_\bot^4 c\lambdabar
\gamma^2}\cr &&\times \int d\vec{r'}_\bot \int d\vec{r''}_\bot
\exp\left[-\frac{r^{'2}_\bot}{2\sigma_\bot^2}\right]
\exp\left[-\frac{r^{''2}_\bot}{2\sigma_\bot^2}\right]
\ln\left(\frac{\left|\vec{r'}_\bot
-\vec{r''}_\bot\right|}{2\lambdabar \gamma}\right) \cr &=& \frac{2
z \gamma_E}{c \lambdabar \gamma^2}+ \frac{2
 z}{ c \lambdabar  {\gamma}^2}  \ln\left(\frac{\sigma_\bot}{{\gamma}
\lambdabar}\right) ~.\cr &&\label{ZIFREE2}
\end{eqnarray}
Also, $Z_{I\mathrm{~free}}$, is logarithmically divergent on
$\sigma_\bot$. This is, in fact, the free-space impedance. The
renormalized impedance\footnote{Note that, without slow-wave
radiative contributions to the field (proportional to $\exp[+i k_w
z]$), it would be impossible to recover Eq. (\ref{IRR2}) and Eq.
(\ref{renim}). In other words, the renormalization process would
fail. This underlines the fact that, although slow-wave radiative
contributions have no realization in the far zone, they are of
fundamental importance in the calculation of the impedance.}, i.e.
the difference of Eq. (\ref{IRR2}) with the free-space impedance
is independent of $\sigma_\bot$ and constitute a result valid for
any value of $K$.

Here we underline the fact that in the limit for $\sigma_\bot
\longrightarrow 0$ the difference between the impedance of an
electron beam moving through a magnetic system and the free-space
impedance is independent of the electron beam model. It is finite,
and can thus be applied in a one-dimensional approximation whereby
the electron bunch is modelled by a line density. This
one-dimensional approach was first proposed in the time-domain to
study Coherent Synchrotron Radiation (CSR) in \cite{SALN} and is
currently used in CSR codes. Also in this case, the renormalized
wake is obtained by subtracting the free-space wake (as is
natural, because the wake is the Fourier transform of the
impedance, i.e. its time-domain counterpart). The
renormalization\footnote{It may be worth to note here, that in the
renormalization procedure used in \cite{SALN} only retarded fields
are used whereas in previous works (see e.g. \cite{DIRA}) devoted
to renormalization in classical electrodynamics a radiation field
is used, that is half the difference of retarded and advanced
fields. } procedure used here and introduced in \cite{SALN} has to
be seen as a mathematical algorithm to deal with calculation of
self-forces of a moving charged line. Such calculation is
problematic due to incompleteness of electromagnetic theory,
yielding to divergence. Such divergence is cancelled by
subtracting the longitudinal force that would be present in a
straight-line motion from the force calculated on a curved
trajectory. The finite difference can be entirely ascribed to
curvature.

\subsection{\label{compwake} Existing studies of the asymptote $\sigma_\bot^2 \ll \lambdabar \lambdabar_w$}

Analytical and numerical studies can be found in literature,
treating longitudinal impedance and wake from an undulator setup
in the case of a line density distribution of electrons (see e.g.
\cite{SALB}, \cite{MART} and \cite{WUST}).

Reference \cite{SALB} deals with the one-dimensional renormalized
wake of an electron beam with a Gaussian longitudinal profile

\begin{eqnarray}
f(s) = \frac{(-e)N}{\sqrt{2\pi} \sigma_z} \exp\left[-\frac{s^2}{2
\sigma_z^2}\right]~, \label{fs}
\end{eqnarray}
$N$ being the number of electrons in the beam. In particular, in
reference \cite{SALB}, the following expression for the energy
gained or lost by a particle at position $z$ down the beamline and
position $s$ within the bunch\footnote{It should be noted that the
definition of $s$ in this paper differs for a sign with respect to
that in \cite{SALB}.} was obtained:

\begin{eqnarray}
\Delta {\mathcal{E}} = \frac{e^2 N K^2 z}{\sqrt{2\pi} \sigma_z
\gamma^2} \bar{G}(p,K,x)~, \label{enlo}
\end{eqnarray}
where $x=-s/\sigma_z$, $p \gg 1$ is the bunch length parameter

\begin{eqnarray}
p = \frac{\gamma^2 k_w \sigma_z}{1+K^2/2} \label{ppp}
\end{eqnarray}
and $\bar{G}$ is given by

\begin{eqnarray}
\bar{G}(p,K,x) = \frac{x}{2} \exp\left[-\frac{x^2}{2}\right]
\left[\ln(p) + g(K)\right] + F(x)~, \label{GGGG}
\end{eqnarray}
with

\begin{eqnarray}
F(x) &=& \frac{1}{4} [\gamma_E + 3 \ln(2) -2]x
\exp\left[-\frac{x^2}{2}\right] \cr && - \sqrt{\frac{\pi}{8}}
\Bigg\{1+\mathrm{erf}\left[\frac{x}{\sqrt{2}}\right] -
x\exp\left[-\frac{x^2}{2}\right] \cr && \times \int_0^x dx'
\exp\left[\frac{(x')^2}{2}\right]
\left[1+\mathrm{erf}\left[\frac{x'}{\sqrt{2}}\right]
\right]\Bigg\}~.\label{Fxdef}
\end{eqnarray}
Moreover, in the limit for $K^2 \ll 1$, $g(K) \longrightarrow 0$,
while in the limit for $K^2 \gg 1$, $g(K) \longrightarrow 1$. For
arbitrary values of $K$, $g(K)$ was presented in \cite{SALB} as a
plot, using numerical integration techniques. Now $g(K)$ can be
expressed fully analytically:

\begin{eqnarray}
g(K) = 1-\frac{\ln\left[1+{K^2}/{2}\right]}{K^2/2}~. \label{gofK}
\end{eqnarray}
Eq. (\ref{enlo}) was already been independently cross-checked,
with the help of the code $\mathrm{TraFiC}^4$, in \cite{MART}. In
this paper we underline the correctness of Eq. (\ref{enlo}).

Reference \cite{WUST} deals with the renormalized impedance in the
case of a line density distribution. Such impedance is presented
in Eq. (26) of \cite{WUST} in the asymptotic case for $K \gg 1$.
When $\sigma_\bot \ll \sqrt{\lambdabar \lambdabar_w}$ and $K^2 \gg
1$ only the first term of Eq. (\ref{IRR2}) survives, and the total
renormalized impedance $Z_{\mathrm{ren}}$ reads:

\begin{eqnarray}
Z_{\mathrm{ren}} = {Z} - i Z_{I\mathrm{~free}} = - \frac{K^2  \pi
z}{4 c \lambdabar \gamma^2} - i \frac{K^2  z}{2 c \lambdabar
\gamma^2}
\ln\left({\frac{\lambdabar}{\lambdabar_r}}\right)~.\label{renim}
\end{eqnarray}
Eq. (\ref{renim}) is in agreement with reference \cite{WUST},
where the impedance per unit length is given\footnote{An extra
factor $-1/c$ in our expression is the result of different
definition of impedance.}.

\subsection{\label{encons} Energy conservation law for $\sigma_\bot^2
\ll \lambdabar \lambdabar_w$}

In general, the real part of the impedance can always be
cross-checked with the energy conservation law, that requires:

\begin{eqnarray}
\frac{d W}{d\omega} = - \frac{1}{\pi}
\left|\bar{f}(\omega)\right|^2 Z_R(\omega) ~, \label{Poy}
\end{eqnarray}
where the energy spectrum of the radiation, $dW/d\omega$, is
defined as the integral over all angles of the total energy
emitted per unit frequency per unit solid angle $d\Omega = \theta
d\theta d\phi$ :

\begin{eqnarray}
\frac{d W}{d \omega} = \int_0^{2\pi} d\phi \int_0^\infty d\theta
\frac{d W}{d\omega d\Omega} = \frac{c z_o^2}{4 \pi^2}
\int_0^{2\pi} d\phi \int_0^\infty d\theta~\theta
\left|\vec{\widetilde{E}}\right|^2 ~.\label{espec}
\end{eqnarray}
It is easy to verify Eq. (\ref{Poy}) in the case $\sigma_\bot^2
\ll \lambdabar \lambdabar_w$. In this case, the electron beam
transverse size is much smaller than the radiation diffraction
size, and a filament-beam model can be used.

Our theory has been developed for any value of the undulator
parameter $K$, and in the long-wavelength asymptote, i.e. for
$\lambdabar_r/\lambdabar \ll 1$. In this case it is possible to
give a simple mathematical description of the radiation energy
spectrum. Based on this expression we will then verify Eq.
(\ref{Poy}).


In order to calculate the energy spectrum according to Eq.
(\ref{espec}) we must first calculate $\vec{\widetilde{E}}_\bot$
in the far zone. We can specify "how near" $\omega$ is to the
resonant frequency $\omega_r=2 k_w c \bar{\gamma}_z^2$ by
introducing a detuning parameter $C$, defined as $C = \omega
/(2\bar{\gamma}_z^2c)-k_w = ({\Delta\omega}/{\omega_r}) k_w$,
where $\omega = \omega_r + \Delta \omega$. Then, the field
generated by a filament-beam is well-known and is given, in
paraxial approximation (see e.g. Eq. (13) of \cite{OURI}), by:

\begin{eqnarray}
&&{\vec{\widetilde{E}}}_\bot= \exp\left[i\frac{\omega \theta^2
z}{2c}\right] \frac{i \omega |\bar{f}(\omega)|}{c^2 z}
\int_{-L_w/2}^{L_w/2} dz'\left\{\frac{K}{2 i
\gamma}\left[\exp\left(2 i k_w z'\right)-1\right]\vec{e}_x
+\vec{\theta}\exp\left(i k_w z'\right)\right\} \cr &&\times
\exp\left[i \left(C + {\omega \theta^2 \over{2 c }}\right) z' -
{K\theta_x\over{\gamma}}{\omega\over{k_w c}}\cos(k_w z')  -
{K^2\over{8\gamma^2}} {\omega\over{k_w c}} \sin(2 k_w z') \right]
~.\label{undurad2}
\end{eqnarray}
As first proposed in \cite{ALFE} one may use the Anger-Jacobi
expansion $\exp\left[i a \sin(\psi)\right] =
\sum_{p=-\infty}^{\infty} J_p(a) \exp\left[ip\psi\right]$, where
$J_p(\cdot)$ indicates the Bessel function of the first kind of
order $p$, to write the integral in Eq. (\ref{undurad2}) in a
different way:

\begin{eqnarray}
&&{\vec{\widetilde{E}}}_\bot= \exp\left[i\frac{\omega \theta^2
z}{2c}\right] \frac{i \omega |\bar{f}(\omega)|}{c^2 z}
\sum_{m,n=-\infty}^\infty J_m(u) J_n(v) \exp\left[\frac{i \pi
n}{2}\right] \cr && \times \int_{-L_w/2}^{L_w/2} dz'\exp\left[i
\left(C + {\omega \theta^2 \over{2 c }}\right) z'\right]
\left\{\frac{K}{2 i \gamma} \left[\exp\left(2 i k_w
z'\right)-1\right]\vec{e}_x +\vec{\theta}\exp\left(i k_w
z'\right)\right\} \cr &&\times \exp\left[i (n+2m) k_w z'\right]
~,\label{undurad3}
\end{eqnarray}
where $u = - {K^2 \omega}/({8 \gamma^2 k_w c})$ and $v = - {K
\theta_x \omega}/({\gamma k_w c})$. There is no simple result
valid at the same time for arbitrary $K$ values and arbitrary
detuning. However, there are asymptotes for
$\lambdabar_r/\lambdabar \ll 1$ and arbitrary $K$, or $K^2 \ll 1$
and arbitrary detuning. We will first assume $K^2 \ll 1$ and
arbitrary values for $\lambdabar / \lambdabar_r$\footnote{Note
that in the long wavelength asymptote, $\lambdabar \gg
\lambdabar_r$, i.e. $\omega/\omega_r \ll 1$, we always have
$|\Delta \omega|/\omega_r \gg 1/N_w$, but not viceversa. Thus,
here we are considering $|\Delta \omega|/\omega_r \gg 1/N_w$, but
arbitrary value for $\lambdabar / \lambdabar_r$.}. Consider the
case $|\Delta \omega|/\omega_r \gg 1/N_w$, $N_w$ being the number
of undulator periods, $C < 0$ and $K^2 \ll 1$. Because of these
assumptions, both $u \ll 1$ and $v \ll 1$ (here $\theta^2 \sim 2 c
|C|/\omega$). This means that asymptotic expansions of Bessel
functions can be used. Non-negligible contributions are for
$n=m=0$ or for $n=-1$ and $m=0$. It follows that

\begin{eqnarray}
{\vec{\widetilde{E}}}_\bot &=& - \frac{\omega |\bar{f}(\omega)|
K}{2 c^2 z \gamma}\exp\left[i\frac{\omega \theta^2 z}{2c}\right]
\left\{\vec{e}_x- \vec{\theta}\frac{ \theta_x \omega}{ k_w
c}\right\} \cr && \times \int_{-L_w/2}^{L_w/2} dz'\exp\left[i
\left(C + {\omega \theta^2 \over{2 c }}\right) z'\right] \cr & =
&- \frac{\omega |\bar{f}(\omega)| K L_w}{2 c^2 z \gamma}
\exp\left[i\frac{\omega \theta^2 z}{2c}\right]
\left\{\left[1-\frac{ \theta_x^2 \omega}{ k_w c}\right]\vec{e}_x
+\left[\frac{ \theta_x \theta_y \omega}{ k_w
c}\right]\vec{e}_y\right\}\cr &&\times
\mathrm{sinc}\left[\frac{L_w}{4}\left(C + {\omega \theta^2 \over{2
c }}\right) \right] ~. \label{undurad4}
\end{eqnarray}
Note that since both $u \ll 1$ and $v \ll 1$, the Anger-Jacobi
expansion is not really necessary here, and we might have derived
Eq. (\ref{undurad4}), based on $K^2 \ll 1$ and $|\Delta
\omega|/\omega_r \gg 1/N_w$, directly from Eq. (\ref{undurad2}) by
directly expanding in Taylor series the exponential function of
trigonometric arguments.

The total energy emitted per unit frequency per unit solid angle
$K^2 \ll 1$ and $|\Delta \omega|/\omega_r \gg 1/N_w$ is

\begin{eqnarray}
\frac{d W}{d\omega d\Omega} = \frac{\omega^2 |\bar{f}(\omega)|^2
K^2 L_w^2}{16\pi^2 c^3 \gamma^2} \left\{\left[1-\frac{ \theta_x^2
\omega}{ k_w c}\right]^2 +\left[\frac{ \theta_x \theta_y \omega}{
k_w c}\right]^2\right\} \mathrm{sinc}^2\left[\frac{L_w}{4}\left(C
+ {\omega \theta^2 \over{2 c }}\right) \right] \cr &&
\label{dWdOdw}
\end{eqnarray}
in agreement with \cite{AMIR}. Substituting Eq. (\ref{dWdOdw}) in
Eq. (\ref{espec}) and using the fact that $N_w \gg 1$ and
$\mathrm{sinc}^2[x/a]/(\pi a) \longrightarrow \delta(x)$ for
$a\longrightarrow 0$ we obtain

\begin{eqnarray}
\frac{d W}{d\omega} = \frac{\omega |\bar{f}(\omega)|^2 K^2 L_w}{8
c^2 \gamma^2} \left[1+\left(\frac{\omega}{c k_w
\bar{\gamma}_z^2}-1\right)^2\right]~, \label{dWdw}
\end{eqnarray}
Also Eq. (\ref{dWdw}) is valid for  $K\ll 1$ and $|\Delta
\omega|/\omega_r \gg 1/N_w$, and is in agreement with \cite{ALFE}
(where the energy spectrum was first calculated) and,
\cite{AMIR}\footnote{A typing error is present in Eq. (2.11) of
\cite{AMIR}.} and \cite{SALB}. In the limit for $\lambdabar \gg
\lambdabar_r$ we write

\begin{eqnarray}
\frac{d W}{d\omega} = \frac{\omega |\bar{f}(\omega)|^2 K^2 L_w}{ 4
c^2 \gamma^2} ~, \label{dWdwlim}
\end{eqnarray}
Eq. (\ref{dWdwlim}) is at the left hand side of Eq. (\ref{Poy})
and has been calculated for $K\ll 1$ and $\lambdabar \gg
\lambdabar_r$.

The right hand side can be written using the real part of Eq.
(\ref{ZRR}), that is valid for arbitrary $K$ values and
$\lambdabar \gg \lambdabar_r$, and calculating the impedance along
an undulator of length $L_w$. We obtain:

\begin{eqnarray}
- \frac{1}{\pi} \left|\bar{f}(\omega)\right|^2 L_w \frac{d
Z_R}{dz}= \frac{\omega  |\bar{f}(\omega)| K^2 L_w}{4 c^2 \gamma^2}
~, \label{poyright}
\end{eqnarray}
thus verifying Eq. (\ref{Poy}) and energy conservation. Agreement
between the asymptote of Eq. (\ref{dWdw}) for $\lambdabar \gg
\lambdabar_r$ and Eq. (\ref{ZRR}) is due to the fact that Eq.
(\ref{ZRR}) is valid for arbitrary values of $K$. However, Eq.
(\ref{dWdwlim}) can also be calculated from Eq. (\ref{undurad2})
under the only assumption $\lambdabar \gg \lambdabar_r$, i.e. Eq.
(\ref{dWdwlim}) is not only valid for $K^2\ll1$, but for any value
of $K$. More generally, we can say that in the long wavelength
asymptote ($\lambdabar \gg \lambdabar_r$) it is sufficient to
account for the first harmonic only, independently of the value of
$K$.

\subsection{\label{sub:aslarg} Asymptotic case for $\sigma_\bot^2 \gg \lambdabar
\lambdabar_w$}

As we already discussed, radiation field and space-charge field
exhibit different formation lengths and different transverse
scales, namely $\sqrt{\lambdabar \lambdabar_w}$ and $\lambdabar
\bar{\gamma}_z$. The same transverse scales are also present in
Eq. (\ref{ZR}) and Eq. (\ref{IR}).

The first two terms in $Y_0$ and $K_0$ in Eq. (\ref{IR}), as well
as the entire real part of the impedance, are linked to the
presence of transverse current density and to radiation field. The
last term in Eq. (\ref{IR}) instead, is due to the presence of
longitudinal space-charge field, a combination of current and
gradient terms. The corresponding Bessel functions yield different
characteristic transverse scales. Bessel functions related with
the radiation field are linked with a transverse size
$\sqrt{\lambdabar \lambdabar_w} \sim \sqrt{\lambdabar \lambdabar_r
\bar{\gamma}_z^2 }$. Those related with the longitudinal
space-charge field are linked with a transverse size $\lambdabar
\bar{\gamma}_z \sim \sqrt{\lambdabar \lambdabar \bar{\gamma}_z^2
}$. Since $\lambdabar \gg \lambdabar_r$, it follows that the
characteristic transverse size related with the radiation field
contribution is much smaller than that related with the
space-charge field contribution, $\sqrt{\lambdabar \lambdabar_w/2}
\ll \lambdabar \bar{\gamma}_z$. By inspection of Eq. (\ref{ZR})
and Eq. (\ref{IR}) one can see that the value of
$\left|\vec{r'}_\bot -\vec{r''}_\bot\right|$ is limited by
$\sigma_\bot$, because of the presence of the exponential
functions under the integration sign. Therefore, assuming constant
total charge of the beam, when the electron beam transverse size
$\sigma_\bot$ increases beyond $\sqrt{\lambdabar \lambdabar_w}$
the radiation contribution is suppressed with respect to the
space-charge one.

Summing up, when condition (\ref{cons2}) is valid together with
(\ref{cons1}), i.e. $\sigma_\bot^2 \gg {\lambdabar \lambdabar_w}$,
we may neglect the real part of the impedance $Z_R$ and
approximate the total impedance with

\begin{eqnarray}
{Z} &=& - i \frac{2 \omega z}{\bar{\gamma}_z^2} \int
d\vec{r'}_\bot \int d\vec{r''}_\bot \rho_o^*(\vec{r'}_\bot)
{\rho_o}\left(\vec{r''}_\bot\right)
K_0\left(\frac{\left|\vec{r'}_\bot
-\vec{r''}_\bot\right|}{\lambdabar \bar{\gamma}_z}\right) ~.
\label{Zapprox}
\end{eqnarray}
This means that, in the limit $\sigma_\bot^2 \gg {\lambdabar
\lambdabar_w}$, the only field to be accounted for when
calculating impedance (and wake), is the effective longitudinal
space-charge field.

\subsection{\label{sub:disc} Discussion}

Results obtained in Section \ref{sub:aslarg}, namely Eq.
(\ref{Zapprox}), mean that in the limit $\sigma_\bot^2 \gg
{\lambdabar \lambdabar_w}$ radiation is suppressed, so that the
beam can be considered as non-radiating, and only space-charge
impedance is present. Such impedance amounts to the free-space
impedance, where $\gamma$ is consistently substituted with
$\bar{\gamma}_z$. Eq. (\ref{Zapprox}) gives the correct impedance
at position $z$ inside the undulator, as an asymptotic limit for
$\sigma_\bot^2 \gg {\lambdabar \lambdabar_w}$ of our general
theory.

Our results are in contrast with \cite{CONF}. Authors of
\cite{CONF} first noted, correctly, that the presence of a finite
transverse dimension of the beam $\sigma_\bot$ suppresses
radiation in the far zone. Thus, the real part of the renormalized
impedance in Eq. (\ref{renim}) (valid in the one-dimensional limit
of a pencil beam) can be generalized to the case when a finite
transverse dimension of the beam $\sigma_\bot$ is present, by
multiplying it by an exponentially suppressing factor. However,
they extended such understanding to the imaginary part as well,
that is incorrect. This led them to obtain the following
expression for the renormalized undulator impedance accounting for
a finite $\sigma_\bot$ (see Eq. (2) of reference \cite{CONF}):

\begin{eqnarray}
Z_{\sigma_\bot} &=& Z_{\mathrm{ren}} \exp\left[- \frac{2\omega}{c}
k_w \sigma_\bot^2 \right] = \left[- \frac{K^2  \pi z}{4 c
\lambdabar \gamma^2} - i \frac{K^2  z}{2 c \lambdabar \gamma^2}
\ln\left({\frac{\lambdabar}{\lambdabar_r}}\right)\right]\exp\left[-
\frac{2\omega}{c} k_w \sigma_\bot^2 \right],\cr &&\label{renim2}
\end{eqnarray}
the explicit expression for $Z_{\mathrm{ren}}$ being given by Eq.
(\ref{renim}).

Eq. (\ref{renim2}) is not derived, in \cite{CONF}, as the
asymptotic result from a complete theory. It is the outcome of an
analogy with the fact that radiation in the far zone is suppressed
when a finite transverse electron beam size $\sigma_\bot$ is
considered. However, the imaginary part of $Z_{\mathrm{ren}}$
results as a combination (a difference, actually) of two
logarithmic contributions, that can be respectively ascribed to
space-charge and radiative field (see Section \ref{sub:smalll})
and present different characteristic transverse scales
($\bar{\gamma}_z \lambdabar$ and $\sqrt{\lambdabar\lambdabar_w}$)
as a consequence of different field formation-lengths.
$Z_{\mathrm{ren}}$ is calculated in the limit for $\sigma_\bot^2
\ll \lambdabar \lambdabar_w$. Only in this limit the dependence on
$\sigma_\bot$ in the imaginary part of $Z_{\mathrm{ren}}$ is
cancelled as a result of the combination of the before-mentioned
logarithmic contributions. For finite transverse size
$\sigma_\bot$ such compensation does not take place at all. For
example, for the ESASE scheme at LCLS (see Section
\ref{sec:expl}), one has $\sigma_\bot = 30 ~\mu$m,
$\sqrt{\lambdabar \lambdabar_w} \simeq 6~\mu$m and $\bar{\gamma}_z
\lambdabar = 500 \mu$m. As a result, $\sigma_\bot$ is small with
respect to $\bar{\gamma}_z \lambdabar$, but large with respect to
$\sqrt{\lambdabar\lambdabar_w}$. The incorrectness of Eq.
(\ref{renim2}) (i.e. Eq. (2) of \cite{CONF}) follows from these
observations.

Authors of reference \cite{CONF} conclude that the impedance in
Eq. (\ref{renim2}), that is related with curved trajectory, is
suppressed. Thus, the space-charge induced, free-space impedance
(Eq. (5) in \cite{CONF}) is finally considered when calculating
the energy spread inside the undulator. This result is
counterintuitive. According to it, when the electron beam does not
radiate ($\sigma_\bot^2 \gg \lambdabar \lambdabar_w$), the
presence of the undulator does not influence the impedance,
independently of the value of the undulator parameter $K$.
However, its incorrectness was not trivial to prove, because it
relies on an apparently correct analogy between real and imaginary
part of the impedance. Only  developing a complete theory the
presence of two separate logarithmic dependencies can be spotted.
Thus, in Eq. (\ref{Zapprox}) we saw, as an asymptotic case of our
theory, that only space-charge impedance is relevant for
$\sigma_\bot^2 \gg \lambdabar \lambdabar_w$, but we additionally
demonstrated that $\gamma$ must be consistently substituted with
$\bar{\gamma}_z$.

Our conclusion is that when $K^2 \gtrsim 1$ the presence of the
undulator strongly influences the longitudinal impedance, whether
the beam radiates or not (i.e. independently of the transverse
size $\sigma_\bot$).

\section{\label{sub:resu} Analytical expression for the
wake function in the steady state case for $\sigma_\bot^2 \gg
\lambdabar \lambdabar_w$}

As discussed in the previous Sections, our derivations hold under
conditions (\ref{cons1}) and drastically simplify under condition
(\ref{cons2}), i.e. for $\sigma_\bot^2 \gg \lambdabar
\lambdabar_w$. In fact, as we have seen before, when condition
(\ref{cons2}) holds transverse contributions to impedance and wake
are negligible. In the following we will consider the case when
both (\ref{cons1}) and (\ref{cons2}) are satisfied, that is the
case for ESASE schemes at LCLS, as we will see in Section
\ref{sec:expl}.

\begin{figure}
\begin{center}
\includegraphics*[width=120mm]{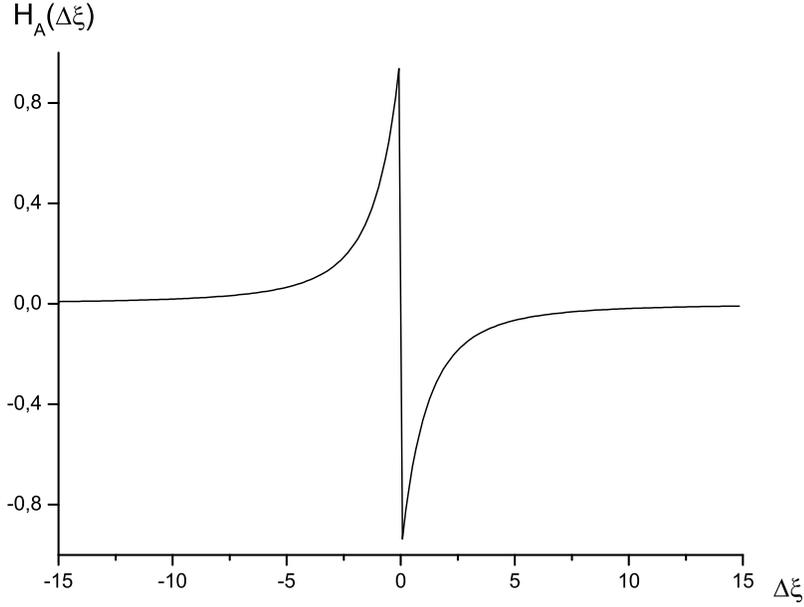}
\caption{\label{ga} Plot of the universal function $H_A$ as a
function of $\Delta \xi$. }
\end{center}
\end{figure}

We will consider a transverse longitudinal profile, as specified
in Eq. (\ref{rhoog}), and a longitudinal bunch profile, specified
by Eq. (\ref{fs}). Note that the $rms$ bunch length $\sigma_z$ is
connected to the $rms$ bunch duration $\sigma_t$ by $\sigma_z =
\beta c \sigma_t$, so that, in terms of time and frequency we have

\begin{eqnarray}
f(t) = \frac{(-e)N}{\sqrt{2\pi} \sigma_t} \exp\left[-\frac{t^2}{2
\sigma_t^2}\right]~~~\longleftrightarrow~~~\bar{f}(\omega) = (-e)N
\exp\left[-\frac{\omega^2 \sigma_t^2}{2}\right]~, \label{ft}
\end{eqnarray}
When $\sigma_\bot^2 \gg \lambdabar \lambdabar_w$, an expression
for the wake can be found by Fourier-transforming the impedance
given in Eq. (\ref{Zapprox})\footnote{In the more general case
when only conditions (\ref{cons1}) hold, we should
Fourier-transform both Eq. (\ref{ZR}) and Eq. (\ref{IR}).}. As
already noted, Eq. (\ref{Zapprox}) is mathematically identical to
the free-space expression where only $\gamma$ has been substituted
by $\bar{\gamma}_z$. Here we will present only the final result
for the wake function. For mathematical details regarding wake
calculations, we refer the interested reader to a previous work of
us \cite{OURL}. That paper dealt with a different subject, namely
wake fields and impedances for electron beams accelerated through
ultra-high field gradients. However, in \cite{OURL}, we also
analyzed the steady-state ($z \gg 2 \bar{\gamma}_z^2 \sigma_z$),
free-space case of Gaussian transverse and longitudinal
distribution for the beam\footnote{In the limit for $z \gg 2
\bar{\gamma}_z^2 \sigma_z$, the antisymmetric part of the
longitudinal wake function $G_A$ (always defined, as in Eq.
(\ref{Gred}), from a source disk to a test disk separated of
$\Delta s$) is dominant with respect to the symmetric part. In
this paper we will only analyze this part, that will be used later
on to discuss the feasibility of ESASE schemes.}. We
find\footnote{See Eq. (21) of reference \cite{OURL}, where
$\gamma$ has been substituted with $\bar{\gamma}_z$.} that the
antisymmetric part of the wake $G_A$ is given by $G_A(\Delta \xi)
= {\bar{\gamma}_z \eta \hat{z}}/{\sigma_\bot}\cdot H_A(\Delta
\xi)$, where

\begin{eqnarray}
H_A(\Delta \xi) =   - \frac{1}{2 \sqrt{\pi}} (\Delta
\xi)~\left\{2\frac{\sqrt{\pi}}{|\Delta \xi|} - \pi \exp
\left[\frac{ (\Delta\xi)^2}{4}\right]
\mathrm{erfc}\left[\frac{|\Delta\xi|}{2}\right]\right\}~.
\label{asym4b}
\end{eqnarray}
Here we defined $\Delta \xi = \bar{\gamma}_z (\Delta
s)/\sigma_\bot$, $\eta = \bar{\gamma}_z \sigma_z/\sigma_\bot$ and
$\hat{z} = {z}/(2\bar{\gamma}_z^2 \sigma_z)$. A plot of the
universal function $H_A$, that is the symmetric part of the wake
in units of $\bar{\gamma}_z \hat{z}/\sigma_\bot$, as a function of
$\Delta \xi$ is given in Fig. \ref{ga}.

The energy change of a single particle at position $s$ within the
bunch due to the reactive part of the wake, averaged over
transverse coordinates is given by the convolution $\Delta
\mathcal{E}_A(s) = (-e) \int_{-\infty}^{\infty} G_A(\Delta s)
f(s-\Delta s) d (\Delta s)$.  An explicit expression for $\Delta
\mathcal{E}_A/\mathcal{E}_o$, with $\mathcal{E}_o = \gamma m_e
c^2$, as a function of $\xi = \bar{\gamma}_z s/\sigma_\bot$:

\begin{eqnarray}
\frac{\Delta \mathcal{E}_A}{\mathcal{E}_o}(\xi) &=&
\frac{I_\mathrm{max}}{ \gamma I_A} ~\eta \hat{z}
\int_{-\infty}^{\infty} d (\Delta \xi) H_A(\xi-\Delta\xi)
\exp\left[-\frac{(\Delta \xi)^2}{2 \eta^2}\right]~.\label{elossa2}
\end{eqnarray}
Note that Eq. (\ref{elossa2}) is a function of $\xi$ but also
depends parametrically on $\eta$, and may be presented as

\begin{eqnarray}
\frac{\Delta
\mathcal{E}_A}{\mathcal{E}_o}\left(\frac{s}{\sigma_z}; \eta\right)
&=& \frac{I_\mathrm{max} \hat{z}}{ \gamma I_A}
F\left(\frac{s}{\sigma_z}; \eta\right)~.\label{elossa3}
\end{eqnarray}
where we indicated the parametric dependence of $\eta$ after the
semicolon and

\begin{figure}
\begin{center}
\includegraphics*[width=120mm]{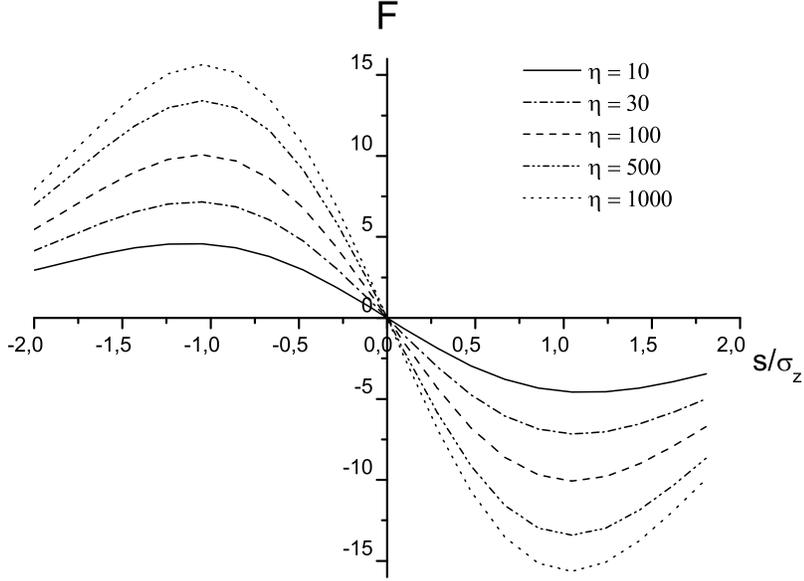}
\caption{\label{fseta} Plot of F in Eq. (\ref{elossaker}) as a
function of $s/\sigma_z$ for different values of $\eta$.}
\end{center}
\end{figure}

\begin{eqnarray}
F\left(\frac{s}{\sigma_z}; \eta\right) &=& \int_{-\infty}^{\infty}
d (\Delta \xi)~ \eta~ H_A\left(\eta
\frac{s}{\sigma_z}-\Delta\xi\right) \exp\left[-\frac{(\Delta
\xi)^2}{2 \eta^2}\right]~.\label{elossaker}
\end{eqnarray}
A plot of $F$ is given as a function of $s/\sigma_z$ in Fig.
\ref{fseta} for different values of $\eta$.

\section{\label{sec:expl} Application to ESASE schemes}

We can now give a practical example of application of our work.
Namely, we calculate the impact of longitudinal wake fields in
ESASE schemes \cite{ZHFA, CONF, ZHPE}.  Here we propose an
analysis on a set of parameters referring to the LCLS \cite{SLAC}
setup considered in \cite{ZHPE}. Similar calculations may be
performed on other parameter sets like those for the European XFEL
\cite{XFEL}.

We consider a beam with normalized emittance after the dispersive
section $\epsilon_n \simeq 1.2$ mm mrad (like in Fig. 3 of
\cite{CONF}). We take the average betatron function in the
focusing lattice $\beta_f = 18$ m, and $\gamma = 2.8 \cdot 10^4$.
This gives a transverse beam size $\sigma_\bot = (\epsilon_n
\beta_f/\gamma)^{1/2} \simeq 30 ~\mu$m. The longitudinal size of
the bunch is $\sigma_z = 50$ nm. The maximal current is about the
Alfven current $I_A \simeq 17$ kA; in fact, $I_\mathrm{peak}
\simeq 18 \mathrm{kA}$. Finally, the undulator has a period
$\lambda_w = 0.03$ m, $K=3.7$, and the vacuum chamber dimension is
$a = 2.5$ mm.

We consider a wavelength $\lambdabar \simeq \sigma_z = 50$ nm. We
can neglect the vacuum chamber influence (see Section
\ref{sec:disc}), because $\bar{\gamma}_z \lambdabar = 500~\mu$m,
as $\bar{\gamma}_z \simeq 10^4$, and $ \bar{\gamma}_z \lambdabar
\ll a =2.5$ mm. The overtaking length is $2 \lambdabar
\bar{\gamma}_z^2 \simeq 10$ m. The saturation length is about $L_s
= 50$ m. Thus $\hat{z} = 5$, according to the definition in
Section \ref{sub:resu} and we can use our asymptotic expression.
Moreover $\eta = \bar{\gamma}_z \sigma_z/\sigma_\bot \simeq 16.7$.
From fig. \ref{fseta} (or from direct calculations) one can see
that the maximal value assumed by $F(s/\sigma_z,\eta)$ for $\eta
=16.7$ is about $F_\mathrm{max} \simeq 6$. It follows that the
energy-chirp peak-to-peak is given by (see Eq. (\ref{elossa2}) or
Eq. (\ref{elossa3})) :

\begin{eqnarray}
{\Delta \mathcal{E}_{A,\mathrm{peak}}} &=& 2 m_e c^2
\frac{I_\mathrm{max}}{I_A}  \hat{z} F_\mathrm{max} \simeq 30
~\mathrm{MeV} ~.\label{elossamax}
\end{eqnarray}
In contrast to this, estimations in \cite{ZHPE} indicate "a swing
in energy of $2.4$ MeV". The reason for this large discrepancy is
due to the fact that in reference \cite{ZHPE}, where it is
correctly recognized that the "most significant cause for concern
is the longitudinal space charge forces", the Lorentz factor
$\gamma$ is incorrectly used in place of $\bar{\gamma}_z$. In
fact, as it is clearly stated in that reference : "While this
expression\footnote{Eq. (3) of reference \cite{ZHPE}.} has been
derived for beam lines containing only drift sections and focusing
elements, we apply it \textit{without modification} to the present
case where the electron beam is passing through the undulator and
oscillates almost rigidly with a deviation of less than $1 \mu$m".

In addition to this, it should be noted that energy chirp is also
accumulated in the free-space between the dispersive section and
the undulator, worsening the situation even more. In the LCLS case
\cite{CONF}, the dispersive section is a dogleg located about
$200$ m from the undulator. One should account for the energy
chirp accumulated in this region too, and sum it to that in Eq.
(\ref{elossamax}). To this extent, reference \cite{OURL} can be
used. The overtaking length is now $2 \lambdabar \gamma^2 \simeq
80$ m, so that $\hat{z} = 2.5$ and our asymptotic expression for
the wake are still valid with some accuracy. Using the same
procedure as for the wake inside the undulator (but considering
$\gamma$ instead of $\bar{\gamma}_z$), we obtain an extra energy
chirp of about ${\Delta \mathcal{E}_{A,\mathrm{peak}}} \simeq 20$
MeV.

The sum of contributions from the straight section after the
dogleg and from the undulator amounts to about $50$ MeV. Although
the energy chirp is non-linear, in order to estimate the magnitude
of the effect we can use the linear energy chirp parameter
$\hat{\alpha}$ defined in \cite{KRIS,SALC}. The effect of linear
energy chirp starts to play a significant role on the FEL gain
when $\hat{\alpha} \gtrsim 1$. Intuitively, this means that the
relative energy change becomes comparable with the FEL parameter
on the scale of the coherence length.  The chirp parameter is
defined as $\hat{\alpha} = -(\gamma \omega \rho_{1D}^2)^{-1} \cdot
{d \gamma}/{dt}$, $\rho_{1D}$ being the one-dimensional
$\rho$-parameter in FEL theory defined as (see \cite{PELL}):

\begin{eqnarray}
\rho_\mathrm{1D} = \frac{\lambda_w}{4\pi} \left[\frac{2 \pi^2 j_o
K^2 A^2_{JJ} }{I_A \lambda_w \gamma^3}\right]^{1/3}
\label{rho1ddef} ~,\end{eqnarray}
where $j_o$ is the beam current density and the coupling factor
$A_{JJ}$, for a planar undulator, is given by $A_{JJ} =
J_0(Q)-J_1(Q)$, where $Q = K^2/(2+K^2)$. For ESASE schemes at LCLS
$I_\mathrm{peak} = 18$ kA, and we have $\rho_{1D}\simeq 10^{-3}$.
Using an estimated peak-to-peak chirp of $50$ MeV we obtain
$\hat{\alpha} \simeq 1$. Thus, the saturation length is
significantly modified \cite{SALC}. This is a reason of concern,
because ESASE schemes are based on the assumption that the nominal
saturation length of about $80$ m is shortened to about $50$ m,
that is only $37.5 \%$ less. The effect described here is
fundamental, in the sense that it cannot be avoided by fine tuning
of the setup parameters.

Finally, it should be noted that in this paper we did not account
for the symmetric part of the wake, related with Transition
Undulator Radiation. This part constitutes only a correction to
our calculations, as the space-charge wake accumulates along the
longitudinal axis, being proportional to $\hat{z}$. We did not
include it in this article, because the space-charge wake alone is
enough to raise concern. Numerical estimations presented in this
paper indicate that effects of energy chirp induces by
space-charge longitudinal wake pose a serious threat to the
operation of ESASE schemes at LCLS.

\section{\label{sec:conc} Conclusions}

In this paper we presented a theory of wake fields in an XFEL
system, with particular emphasis to ESASE schemes \cite{ZHFA,
CONF,ZHPE}.

We worked with specific constraints on parameters, that are
fulfilled in XFEL setups (see Section \ref{sec:disc}). Namely, we
neglected the influence of the vacuum chamber and we assumed that
the saturation length is long with respect to the overtaking
length. Our results are valid for arbitrary values of the
undulator parameter $K$ and in the long wavelength asymptotic,
i.e. $\lambdabar \gg \lambdabar_r$, $\lambdabar_r$ being the
reduced wavelength of the fundamental harmonic. Note that, for any
FEL setup, the lasing part of the bunch is always much longer than
$\lambdabar_r$ so that condition $\lambdabar \gg \lambdabar_r$ is
very natural. It follows that our results are of practical
importance not only in relation with ESASE schemes, but for any
FEL setup. We derived expressions for the steady state impedance,
that is composed of a radiative and a space-charge part. Radiation
field and space-charge field are characterized by different
formation lengths:  the undulator period $\lambdabar_w$ and the
overtaking length $2 \lambdabar \bar{\gamma}_z^2$, respectively.
As a result, the steady state radiative part of the impedance can
be applied for any undulator system (with $N_w \gg 1$), whereas
the steady state space-charge part of the impedance can be used
only assuming that the saturation length is long with respect to
the overtaking length, which limits its practical region of
applicability. Non-steady state results for the space-charge part
of the impedance can be obtained applying methods presented in
\cite{OURL} and \cite{BOSH}.

After having dealt with a generic expression for the steady-state
impedance, we specialized our theory to the case when the
transverse beam size $\sigma_\bot^2 \gg \lambdabar \lambdabar_w$.
Major simplifications arise in this case: in particular,
space-charge contributions to impedance and wake dominate with
respect to radiative contributions. In this particular condition,
that is practically fulfilled for ESASE XFEL setups, we showed
that the (antisymmetric) wake can be given in terms of an
asymptotic expression for the wake generated by a beam in uniform
motion along the longitudinal axis (see \cite{OURL}), provided
that the Lorentz factor $\gamma$ is consistently substituted with
the average longitudinal Lorentz factor $\bar{\gamma}_z$. Final
expressions are presented in the case of a planar undulator.
However, there are no specific effects related with such choice,
and our work may be straightforwardly extended to the case of a
helical undulator as well.

We applied our theory to calculate the effects of longitudinal
wake fields on ESASE schemes. Our conclusion is that longitudinal
wake fields pose a threat to the practical realization of ESASE
schemes. This finding is in contrast with estimations in
literature, where no important detrimental effect is foreseen. The
reason for this contrast is an incorrect application, in
literature, of expressions that are valid for beam lines
containing drift sections and focusing elements to describe the
case of XFEL undulators, where the longitudinal Lorentz factor is
sensibly different.

\section*{\label{sec:graz} Acknowledgements}

We thank Martin Dohlus (DESY) for many useful discussions, Massimo
Altarelli and Jochen Schneider (DESY) for their interest in this
work.

\end{document}